\documentclass[preprint,12pt]{elsarticle}
\usepackage[utf8]{inputenc}
\usepackage{times}
\usepackage{amsthm,amsmath,amsfonts,latexsym,amssymb}
\usepackage{graphics,fancyhdr,graphicx}
\usepackage{multirow}
\usepackage{float}
\usepackage{fleqn}
\usepackage{subfigure}
\usepackage{verbatim}
\usepackage{rotating}
\usepackage{makecell}
\usepackage{bm}
\usepackage{color}
\usepackage{bm}
\def \hb{\mathbf{h}}
\def \cb{\mathbf{c}}
\usepackage{colortbl}
\usepackage[table]{xcolor}
\newcommand{\mcL}{\mathcal{L}}

\usepackage{geometry}
\usepackage{setspace}
\biboptions{numbers,sort&compress}
\definecolor{listinggray}{gray}{0.9}
\definecolor{lbcolor}{rgb}{0.9,0.9,0.9}
\definecolor{Darkgreen}{RGB}{0,100,0}

\begin{document}
\abovedisplayskip=6.0pt
\belowdisplayskip=6.0pt
\begin{frontmatter}

\title{Deep neural operators can predict the real-time response of floating offshore structures under irregular waves}

\author[1,2]{Qianying Cao}
\ead{qianying_cao@brown.edu}
\author[2]{Somdatta Goswami}
\ead{somdatta_goswami@brown.edu}
\author[5]{Tapas Tripura}
\ead{tapas.t@am.iitd.ac.in}
\author[5,6]{Souvik Chakraborty}
\ead{souvik@am.iitd.ac.in}
\author[2,3,4]{George Em Karniadakis\corref{cor1}}
\ead{george_karniadakis@brown.edu}
\address[1]{State Key Lab of Coastal and Offshore Engineering, Dalian University of Technology}
\address[2]{Division of Applied Mathematics, Brown University}
\address[3]{School of Engineering, Brown University}
\address[4]{Computational Math Group, Pacific Northwest National Laboratory}
\address[5]{Department of Applied Mechanics, Indian Institute of Technology Delhi}
\address[6]{School of Artificial Intelligence, Indian Institute of Technology Delhi}
\cortext[cor1]{Corresponding author.}

\begin{abstract}
The utilization of neural operators in a digital twin model of an offshore floating structure holds the potential for a significant shift in the prediction of structural responses and health monitoring, offering valuable real-time control insights. In this work, we investigate the effectiveness of three neural operators, namely the deep operator network (DeepONet), the Fourier neural operator (FNO), and the Wavelet neural operator (WNO), to accurately capture the responses of a floating structure under six different sea state codes $(3-8)$ based on the wave characteristics described by the World Meteorological Organization (WMO). To further enhance the accuracy of the vanilla architecture of the neural operators, novel extensions, such as wavelet-DeepONet and self-adaptive WNO, are proposed in this paper. The results demonstrate that these high-precision neural operators can deliver structural responses more efficiently, up to two orders of magnitude faster than a dynamic analysis using conventional numerical solvers. Additionally, compared to gated recurrent units (GRUs), a commonly used recurrent neural network for time-series estimation, neural operators are both more accurate and efficient, especially in situations with limited data availability. Taken together, our study shows that FNO outperforms all other operators for approximating the mapping of one input functional space to the output space as well as for responses that have small bandwidth of the frequency spectrum. Conversely, DeepONet, with historical states, proves most accurate in learning the mapping of multiple input functions to the output space and capturing responses within a broad frequency spectrum.
\end{abstract}


\end{frontmatter}

\section{Introduction}

Deep-water explorations are increasingly being undertaken in harsh environments to meet the growing demands for resources and energy. As a result, there is a critical need for offshore floating structures that are both safe and reliable, as structural failures in deep water can lead to substantial economic and environmental consequences. These floating structures can move in six degrees of freedom (DOFs) when subjected to environmental forces, but their motion is constrained by mooring systems. Predicting the response of these structures to incident waves presents a significant challenge due to the complex characteristics involved, such as nonlinearity, wide-band spectrum, and multi-scale features. For instance, the surge motion of a semi-submersible platform consists of wave frequency and low-frequency components, each with different timescales. Understanding and accurately predicting these responses are crucial given the aforementioned constraints.

In recent years, data-driven machine learning algorithms, including dynamic mode decomposition~\cite{diez2022time}, sparse identification for nonlinear dynamics (SINDy)~\cite{fukami2021sparse}, and deep neural networks (DNNs)~\cite{goswami2022physics,guo2021predicting}, have emerged as viable methodologies for addressing the challenges associated with accurate predictions of complex responses for offshore floating structures~\cite{guo2021predicting,ye2022emd,schirmann2022data,liong2022data}. These methodologies have led to advancements in neural simulators~\cite{christiansen2013efficient,sidarta2017prediction,lee2022standardization,arifuzzaman2022nonlinear,cotrim2022neural,kwon2022prediction,qiao2021realtime,wang2022new}, which simulate structural responses to environmental excitation. Real-time response prediction, which involves forecasting future responses based on past excitation or responses, has been achieved using DNNs. Notable examples include the prediction of heave and surge motions of a semi-submersible platform and short-term roll and sway predictions using hybrid empirical mode decomposition (EMD) models~\cite{ye2022emd}. Some papers also employ machine learning methods in predicting nonlinear and complex dynamics of offshore vessels and structures~\cite{del2021learning,d2021recurrent,guo2022real,guth2022wave,guth2023statistical}.
Del Aguila Ferrandis~\cite{del2021learning} trained the recurrent neural networks of LSTM type to forecast nonlinear motions in irregular long-crested head and oblique seas. Furthermore, in another work, D' Agostino et al.~\cite{d2021recurrent} investigated the prediction capability of recurrent-type neural networks for real-time prediction of ship motions in high sea state. Guo et al.~\cite{guo2022real} performed real-time prediction of $6$ degrees of freedom motions of a turret-moored FPSO in a harsh sea state based on the GRU model. Guth and Sapsis~\cite{guth2022wave} developed a wave episode based Gaussian process regression framework to model the non-Gaussian statistics of ship loads due to nonlinear interactions with irregular waves, which only needs a fraction of the data and computational time compared with Monte Carlo simulations. While these applications based on data correlation have shown merit in motion prediction, they often do not incorporate physics-based information. Physics-informed neural network (PINN) which embeds the physics into neural operator, have shown to be effective tool in many fields~\cite{wu2023comprehensive,anagnostopoulos2023residual,zhang2023artificial}. Li et al.~\cite{lu2023motion} explored the use of PINNs to estimated motion and identify system parameters of the moored buoy under different sea states.

Neural simulators, which aim to efficiently simulate structural reactions to environmental excitation, have mainly focused on developing neural solvers, mapping specific systems and learning the mapping between finite-dimensional Euclidean spaces. However, for efficient surrogate models, higher levels of abstraction are required to generalize the surrogate model for unseen input signals. A new line of work has emerged proposing the learning between infinite-dimensional spaces using continuous nonlinear operators developed with DNNs. Neural operators construct interpretable mappings between functions, enabling them to learn entire families of equations (parametrized partial differential equations) and exhibit enhanced generalization ability. These operators are designed to learn the mappings from data that can be either generated using a numerical solver and transfer learned on a real dataset \cite{goswami2022deep}, and/or from historical data \cite{meng2023variational}. The deep neural operator (DeepONet), proposed by Lu et al.~\cite{lu2021learning}, accurately and efficiently learns nonlinear continuous operators, while the Fourier Neural Operator (FNO)~\cite{li2020fourier} and the Wavelet Neural Operator (WNO)~\cite{tripura2022wavelet} extend the idea of neural operators to the frequency domain.

This paper aims to leverage these three main classes of neural operators to accurately learn the underlying family of differential equations governing the responses of a moored semi-submersible platform. The accuracy of the neural operators' solutions for unseen cases will be compared against the conventional gated recurrent unit (GRU). The study considers wave elevation as the input and obtains responses such as platform motions and top tensions of mooring lines as outputs. Separate neural operators are considered for scenarios with zero and non-zero initial conditions, as the response solutions differ significantly between these cases. The paper also explores the operators' ability to learn from limited labeled datasets and introduces two novel extensions, namely the wavelet-DeepONet (W-DeepONet) and the self-adaptive WNO (SA-WNO), which have demonstrated improved prediction accuracy compared to their vanilla counterparts.

The remainder of the paper is organized as follows: Section \ref{sec:problem} presents the numerical solution for computing the response of a floating structure under incident waves and describes the data generation process. Section \ref{sec:methods} details the architectures of the neural operators, including the proposed extensions (W-DeepONet and SA-WNO). Section \ref{sec:results} compares the performance of the studied models for the responses of a floating structure. Finally, Section \ref{sec:summary_and_discussions} summarizes the observations and provides concluding remarks.

\section{Formulation for computing the response of a floating structure}
\label{sec:problem}

The problem of computing the response of a floating structure under incident waves involves two nonlinear operators, \textit{i.e.}, the hydrodynamic operator and the structural operator. The hydrodynamic operator can be regarded as the incompressible Navier-Stokes equation, which takes as input the wave elevation $\eta(t)$ and outputs the hydrodynamic coefficients (i.e., the added mass and potential damping) and wave forces. The structural operator is defined through the equation of motion of the floating structure, which takes the hydrodynamic coefficients and wave force as inputs and yields the structural response as output.

In this study, we aim to learn a combined operator that integrates both the hydrodynamic and structural operators. We consider the wave elevation as the input and obtain the structural responses as outputs. These responses include the motion of the floater characterized by six degrees of freedom and the top tensions of mooring lines. The combined operator for a moored rigid floating structure under incident waves can be represented by the following equation~\cite{cummins1962impulse}:
\begin{align}\label{eq:MCK}
  (\mathbf{M}+\mathbf{A})\ddot{\mathbf{x}}(t)+\int_0^t\mathbf{K}(t-\tau)\dot{\mathbf{x}}(\tau)d\tau+\mathbf{C}\mathbf{x}(t)+\mathbf{f}_{moor}(\mathbf{x}(t),\dot{\mathbf{x}}(t))=\mathbf{f}_{wa}(\eta(t)),
\end{align}
where $\mathbf{M}, \mathbf{A}, \mathbf{C} \in \mathbb{R}^{6\times 6}$ are the body mass, constant added mass, and the linearized hydrostatic restoring force coefficient, respectively, $\mathbf{K}(t)$ is the retardation function, which relies on the geometry of the floating structure, and $\mathbf{\ddot{x}}(t), \mathbf{\dot{x}}(t), \mathbf{x}(t)$ $\in \mathbb{R}^{6\times 1}$ are the acceleration, velocity, and displacement of the floating structure, respectively. The mooring force $\mathbf{f}_{moor}(\mathbf{x}(t),\dot{\mathbf{x}}(t)) \in \mathbb{R}^{6\times 1}$ is a nonlinear function, which is related to the motions of the platform. The wave force $\mathbf{f}_{wa}(\eta(t)) \in \mathbb{R}^{6\times 1}$ acting on the floater is caused by the incident wave elevation $\eta(t)$, which can be calculated by a diffraction analysis~\cite{faltinsen1993sea,pinkster1979mean}. Typically, for a given wave elevation, $\eta(t)$, we obtain $\mathbf{f}_{wa}$, and then using the computed $\mathbf{f}_{wa}$, we solve Eq.~\ref{eq:MCK} for computing the displacement $\mathbf{x}(t)$.

\noindent The learned operator, $\mathbb G_{\mathbf{x}}$, of the structural response, that is the solution of Eq.~\ref{eq:MCK}, can be written as
\begin{align}\label{eq:operator_platform}
  \mathbf{x}(t)=\mathbb G_{\mathbf{x}}(t;\mathbf{x}(t_0),\dot{\mathbf{x}}(t_0),\eta(t)), 
\end{align}
where $\mathbf{x}\in \mathbb{R}^{6\times 1}$, and $\mathbf{x}(t_0)$ and $\dot{\mathbf{x}}(t_0)$ denote the initial displacements and velocities of the six DOFs, respectively. The six DOFs are decoupled as shown in Eq.~\ref{eq:operator_platform} and are only related to the wave elevation and the initial motions of the six-degrees-of-freedom.
Thus, Eq.~\ref{eq:operator_platform} can be expanded as:
\begin{align}\label{eq:operator_platform1}
  x_i(t) = \mathbb G_{x_i}(t;x_1(t_0),\dot{x}_1(t_0),x_2(t_0),\dot{x}_2(t_0),\cdots, x_6(t_0),\dot{x}_6(t_0),\eta(t)),
\end{align}
where $x_i \in \mathbf{x}$ and $i = \{1, 2, \cdots, 6\}$. The top tensions of the mooring lines have a direct relationship with the motion of the platform, so the operator of the top tension, $\mathbb G_{\mathbf{T}}$, has the following form:
\begin{align}\label{eq:operator_mooring}
  T_{moor}(t)=\mathbb G_{\mathbf{T}}(t;\mathbf{x}(t_0),\eta(t)).
\end{align}
When the initial conditions are kept constant, Eqs.~\ref{eq:operator_platform} and \ref{eq:operator_mooring} can be simplified as:
\begin{subequations}\label{eq:operators}
\begin{align}\label{eq:operator_platform2}
  \mathbf{x}(t)=\mathbb G_{\mathbf{x}}(t;\eta(t)),
\end{align}
\begin{align}\label{eq:operator_mooring1}
  T_{moor}(t)=\mathbb G_{\mathbf{T}}(t;\eta(t)).
\end{align}
\end{subequations}
Eqs.~\ref{eq:operator_platform1}-\ref{eq:operator_mooring1} illustrate that separate neural operators will be acquired for each response.

To learn the mappings mentioned above using neural networks, it is necessary to have a labelled dataset of wave elevations and the corresponding structural responses, which can be obtained through various methods such as numerical simulations, physical experiments, or field tests. Numerical software packages like Orcaflex, Sesam, and ANSYS have been developed and widely utilized for computing structural responses to wave elevations.
In this example, we will use Orcaflex software to illustrate how to obtain the required data by solving Eq.~\ref{eq:MCK}. Fig. \ref{semi-submersible} depicts the configuration of a semi-submersible platform. To obtain the data, Orcaflex requires inputs such as the hydrodynamic coefficients of the platform, the structural model, and the wave spectrum. In Orcaflex, the platform is treated as a rigid body with six degrees of freedom. The hydrodynamic coefficients of the platform can be determined by applying diffraction analysis, which involves solving potential theory problems. Once the wave spectrum is selected, irregular wave elevations can be generated using the linear superposition technique of linear wave components~\cite{hu2018laplace}:
\begin{align}\label{eq:simu}
    \eta(t_j) = \sum_{n=1}^{N} A_n \cos(\omega_n t_j + \theta_n), ~~~~~~~~~~j=0,1,\cdots,N-1, 
\end{align}
where $A_n = \sqrt{2S(\omega_n) \Delta \omega}$, in which $S(\omega_n)$ is the wave spectrum parameterized by the significant wave height and zero-crossing wave period, and $\theta_n$ are randomly generated phase angles uniformly distributed between $0$ and $2 \pi$. The corresponding structural responses are obtained by solving Eq.~\ref{eq:MCK} by the Orcaflex based on the displacement RAOs (response amplitude operators). The reader is referred to Ref.~\cite{orcina2018orcaflex} for
details about the Orcaflex.
The time length studied for each sample is $200$s with a time interval, $\Delta t=0.2$s, hence the total number of temporal points $n_t = 1000$. For generating labeled data to learn the operators in Eqs.~\ref{eq:operator_platform1} and \ref{eq:operator_mooring1} with zero initial conditions, Eq.~\ref{eq:MCK} is solved using the numerical solvers. However, for learning the operators in Eqs.~\ref{eq:operator_platform} and \ref{eq:operator_mooring} with different initial conditions, we first use the solver to obtain the structural responses with zero initial conditions. Then, the wave elevations and the corresponding responses, truncated from non-zero time instants (e.g., $400$ seconds), are used as the training datasets. For example, we truncate the wave elevations and corresponding responses from $400$s$-600$s (total temporal points = $1000$), which are used as training/testing data. In this way the initial conditions (including initial displacement and velocity) for different samples are different.

In the next section, we elaborate on the architectures of the neural operators, the deep operator network (DeepONet), the Fourier Neural Operator (FNO), and the Wavelet Neural Operator (WNO). Additionally, we also introduce useful extensions of these operators, especially the two new proposed extensions-wavelet-DeepONet and self-adaptive WNO.

\section{Neural Operators}
\label{sec:methods}

\begin{figure}[H]
\centering
\includegraphics[width=\textwidth, trim = 0cm 7.2cm 23cm 0cm, clip]{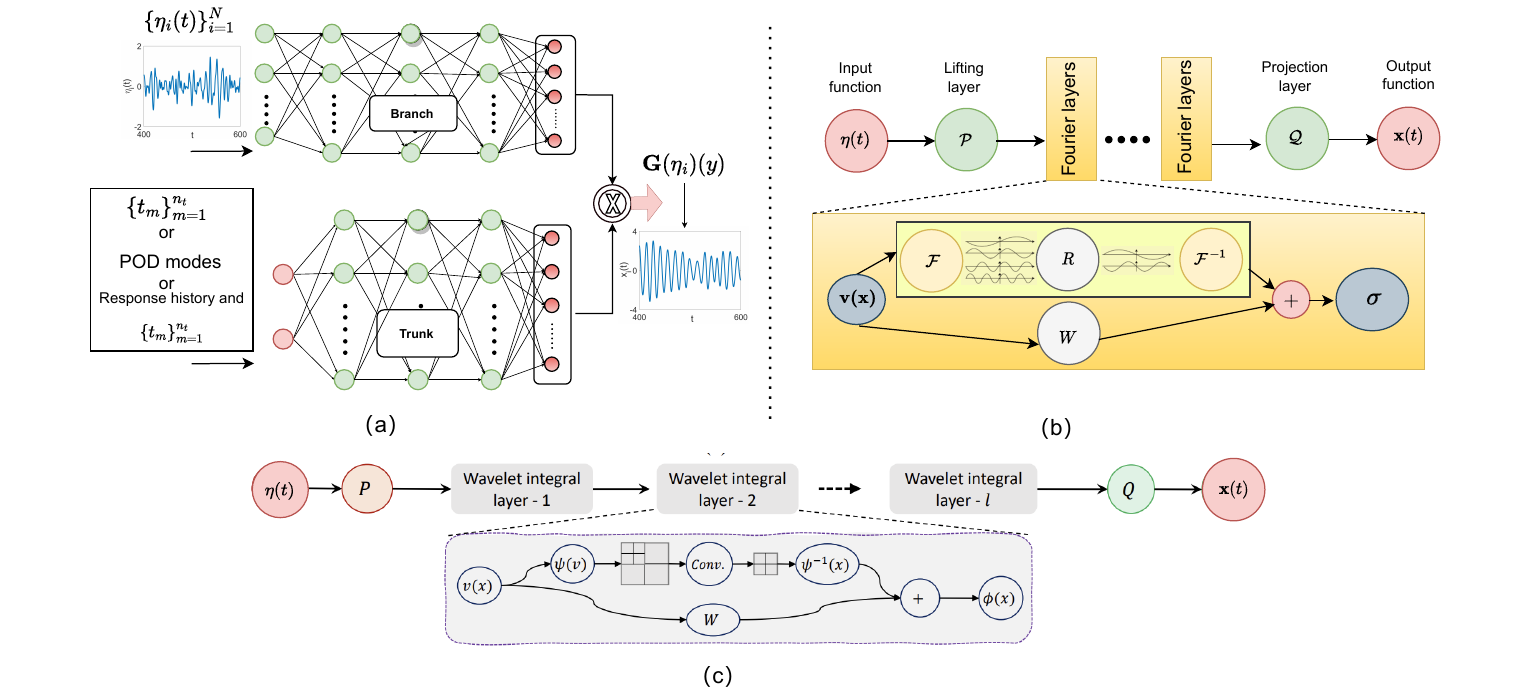}\\
\caption{Architectures of the three neural operators: (a) Deep operator network. If the input to the trunk net is temporal co-ordinates, $\{t_m\}_{m=1}^{n_t}$, it is vanilla DeepONet; if the POD modes are considered in the trunk net, then it is POD-DeepONet; and when the inputs to the trunk net are the historical states along with temporal co-ordinates, it is DeepONet with history), (b) Fourier neural operator, and (c) Wavelet neural operator. For scenarios with zero initial conditions, the branch net of vanilla DeepONet takes as input the wave elevations, and for varying initial conditions, we introduce a second branch network to input the additional initial conditions. In the Fourier and the Wavelet neural operators, the input to the network are wave elevations defined on an equally spaced temporal grid (for zero initial condition). Additionally, for varying initial conditions, the conditions are repeated $n_t$ times to concatenate with the wave elevation and are considered as inputs to the operators.}\label{Figure_neural_operator}
\end{figure}

Neural operators learn nonlinear mappings between infinite dimensional functional spaces and provide a unique simulation framework for predicting complex dynamics~\cite{goswami2022physics}. Once the neural operators are trained, they can generalize for unseen cases, which means the same network parameters are shared across different input functions. We have implemented three operator networks that have shown promising results so far, the DeepONet \cite{lu2021learning}, the Fourier neural operator (FNO) \cite{li2020fourier}, and the Wavelet neural operator (WNO) \cite{tripura2022wavelet}. Although the original DeepONet architecture proposed in \cite{lu2021learning} has shown remarkable success, several extensions have been proposed in \cite{lu2022comprehensive,goswami2022neural,goswami2022deep} to modify its implementation and produce efficient and robust architectures. The architectures of DeepONet, FNO, and WNO are shown in Fig.~\ref{Figure_neural_operator}.

\subsection{Deep Operator Network}

The Deep Operator Network (DeepONet) is based on the concept of the universal approximation theorem for operators~\cite{chen1995universal}. The DeepONet architecture comprises two Deep Neural Networks (DNNs): the branch net, responsible for encoding the input function $\eta(t)$ at specific sensor points, and the trunk net, which encodes information about the temporal coordinates $t_i$ where the solution operator is evaluated to calculate the loss function. These coordinates range from $i=1$ to $i=1000$. To gain a better understanding of the mathematical basis of Deep Neural Networks, we examine a network with $L$ hidden layers. The input layer is denoted as the $0$-th layer, and the output layer is denoted as the $(L+1)$-th layer. The weighted input $\bm z^l_i$ to the $i$\textsuperscript{th} neuron on layer $l$ is determined by the weight $\bm W^l_{ij}$, the bias $\bm b^{l-1}_j$, and can be represented as:
\begin{align} \label{eq:weighted_input}
    \bm z^l_i =  \mathcal{R}_{l-1}\left(\sum_{j=1}^{m_{l-1}}\left(\bm W^l_{ij}(\bm z^{l-1}_j) + \bm b^{l}_j\right)\right),
\end{align} \smallskip
\noindent where $\mathcal{R}_{l-1}\left( \cdot \right)$ denotes the activation function of layer $l$, and $m_{l-1}$ is the number of neurons in layer $l-1$.
Based on the above concepts, the feed-forward algorithm for computing the output $\bm{Y}^L$ is expressed as follows:
\begin{align}\label{eq:feedforward}
    \begin{split}
       \bm{Y}^L &= \mathcal{R}_L(\bm W^{L+1}\bm{z}^L + \bm b^L)\\
       \bm{z}^L & = \mathcal{R}_{L-1}\left(\bm{W}^{L}\bm{z}^{L-1} + \bm b^L\right)\\
       \bm{z}^{L-1} & = \mathcal{R}_{L-2}\left(\bm{W}^{L-1}\bm{z}^{L-2} + \bm b^{L-1}\right)\\
        &\;\;\;\;\;\;\;\vdots\\
        \bm z^1 &= \mathcal{R}_0\left(\bm W^1\bm{x} + \bm b^1\right),
    \end{split}
\end{align}
where $\bm{x}$ is the input of the neural network. Eq.~\ref{eq:feedforward} can be encoded in compressed form as $\bm Y = \mathbb N (x;\bm{\theta})$, where $\bm{\theta} = \left(\bm W, \bm b \right)$ includes both the weights and biases of the neural network $\mathbb N$. 

\noindent The branch network receives input functions representing realizations denoted as $\bm{\eta} = {\bm{\eta}_1, \bm{\eta}_2, \ldots, \bm{\eta}_N}$ for $N$ samples. These functions are discretized as $\bm{\eta}_i = {\eta_i(t_1), \eta_i(t_2), \ldots, \eta_i(t_{n{sen}})}$, where $n{sen}$ is the number of sensor locations and 
$i$ ranges from $1$ to $N$. In the case of a time series problem, $n{sen}$ is equal to the number of time points, denoted as $n_t$. The trunk network takes as input the temporal locations $\bm y = {t_1,t_2,\cdots,t_{n_t}}$ to evaluate the solution operator. Assuming the branch network has $l_{br}$ hidden layers, with the $(l_{br}+1)$\textsuperscript{th} layer being the output layer containing $q$ neurons, when an input function $\bm \eta_{i}$ is given to the branch network, it produces an output feature represented as $[br_1, br_2, \ldots, br_q]^\mathrm{T}$. The output $\bm{z}_{br}^{l{br}+1}$ of the feed-forward branch neural network can be expressed as:
\begin{align} \label{eq:output_branch}
    \begin{split}
    \bm{z}_{br}^{l_{br}+1} &= \left[br_1, br_2, \ldots, br_q\right]^\mathrm{T}\\
    &=\mathcal{R}_{br}\left(\bm W^{l_{br}}\bm{z}^{l_{br}} + \bm b^{l_{br}+1}\right),
    \end{split}
\end{align}
where $\mathcal{R}_{br}\left( \cdot \right)$ denotes the nonlinear activation function for the branch net and $$\bm{z}^{l_{br}} = f_{br}(\eta_i(t_1), \eta_i(t_2), \ldots, \eta_i(t_{n_{sen}})),$$ where $f_{br}\left(\cdot\right)$ denotes a branch net function. Similarly, consider a trunk network with $l_{tr}$ hidden layers, where the $(l_{tr}+1)$-th layer is the output layer consisting of $q$ neurons. The trunk net outputs a feature embedding $[tr_1, tr_2, \ldots, tr_q]^\mathrm{T}$. The output of the trunk network can be represented as:
\begin{align} \label{eq:output_trunk}
    \begin{split}
    \bm{z}_{tr}^{l_{tr}+1} &= \left[tr_1, tr_2, \ldots, tr_q\right]^\mathrm{T}\\
    &=\mathcal{R}_{tr}\left(\bm W^{l_{tr}}\bm{z}^{l_{tr}} + \bm b^{l_{tr}+1}\right),
    \end{split}
\end{align}
where $\mathcal{R}_{tr}\left( \cdot \right)$ denotes the nonlinear activation function for the trunk net and $\bm{z}^{l_{tr}-1} = f_{tr}(t_1, t_2, \ldots, t_{n_t})$. The key point is that we uncover a new operator $\mathbb G_{\bm{\theta}}$ (which is a generalized representation of $\mathbb G_{\mathbf{x}}$ and $\mathbb G_{\mathbf{T}}$) as a neural network that can infer quantities of interest from unseen and noisy inputs. The two networks are trained to learn the solution operator such that
\begin{align}
    \mathbb G_{\bm{\theta}}:\bm{\eta}_i \rightarrow \mathbb G_{\bm{\theta}}(\bm{\eta}_i),\;\; \forall\;\; i = \{1,2,3, \ldots, N\}.
\end{align} 
For a single input function $\bm \eta_i$, the DeepONet prediction $\mathbb G_{\bm \theta}(\bm \eta)$ evaluated at any coordinate $\bm y$ can be expressed as:
\begin{align} \label{eq:output_deeponets}
    \begin{split}
    \mathbb G_{\bm{\theta}}(\bm \eta_{i})(\bm y) &= \sum_{k = 1}^{q}\left(\mathcal{R}_{br}(\bm W^{l_{br}}_k\bm{z}^{l_{br}-1}_k + \bm b^{l_{br}}_k)\cdot \mathcal{R}_{tr}(\bm W^{l_{tr}}_k\bm{z}^{l_{tr}-1}_k + \bm b^{l_{tr}}_k)\right)\\
    &= \sum_{k = 1}^{q}b_k(\eta_i(t_1), \eta_i(t_2), \ldots, \eta_i(t_{n_{t}}))\cdot t_k(\bm y).
    \end{split}
\end{align} 
DeepONet relies on large datasets that contain paired input-output observations. However, it offers a straightforward and intuitive model architecture that allows for fast training. This architecture enables the representation of target output functions in a continuous manner, independent of resolution. Additionally, in a related study by Kontolati et al.~\cite{kontolati2022influence}, they demonstrated the stability of DeepONet's predictions when tested with noisy data. Traditionally, the trainable parameters of DeepONet, denoted as $\bm{\theta}$ in Eq.~\ref{eq:output_deeponets}, are obtained by minimizing a loss function. Common loss functions used in the literature include the $L_1$- and $L_2$-loss functions, defined as:
\begin{align} \label{eq:L1_loss}
\begin{split}
    \mathcal L_1 &= \sum_{i =1}^n \sum_{j =1}^p \big| \mathbb G(\bm \eta_{i})(\bm y_j) - \mathbb G_{\bm{\theta}}(\bm \eta_{i})(\bm y_j)\big|\\
    \mathcal L_2 &= \sum_{i =1}^n \sum_{j =1}^p\big(\mathbb G(\bm \eta_{i})(\bm y_j) - \mathbb G_{\bm{\theta}}(\bm \eta_{i})(\bm y_j)\big)^2,\\
\end{split}
\end{align} 
where $\mathbb{G}_{\bm{\theta}}(\bm \eta_{i})(\bm y_j)$ is the predicted value obtained from the DeepONet, and $\mathbb G(\bm \eta_{i})(\bm y_j)$ is the target value. In all the experiments carried out in this work, we have considered a feed-forward fully connected neural network for the branch and the trunk networks.
\bigbreak
\noindent
\textbf{POD-DeepONet}\\
This approach, presented by Lu et al.~\cite{lu2022comprehensive}, is an extension of the standard DeepONet discussed earlier. In the standard DeepONet, the trunk net is responsible for learning the basis of the output function from the available data. However, in this extended approach, the basis functions are computed in advance using proper orthogonal decomposition (POD) applied to the labeled output of the training data (with the mean value excluded). The labeled outputs are represented as $\mathbb G(\bm \eta_{i})(\bm y)$, where $\bm y$ represents the coordinates at which the outputs are calculated. The POD basis is then utilized in the trunk net. In the branch net, a Deep Neural Network (DNN) is used to learn the coefficients associated with the POD basis, enabling the expression of the output as:
\begin{align}
    \mathbb G_{\bm{\theta}}(\bm \eta_{i})(\bm y) = \sum_{k=1}^q br_k(\eta) \phi_k(t) + \phi_0(t),
\end{align}
where $\phi_0(t)$ is the mean function of all $\bm \eta_i(\bm y), k =1,\dots,q$ computed from the training dataset, and $\{\phi_1, \phi_2, \dots, \phi_q\}$ are the $q$ precomputed POD modes of $\mathbb G(\bm \eta_{i})(\bm y) $. 
\bigbreak
\noindent
\textbf{DeepONet with historical information}\\
This is an adapted version of the DeepONet architecture that includes additional features related to the solution in the trunk network. In this modified architecture, the trunk network receives the historical states of the time signal, denoted as $n_h$ historical features, along with the time step as input~\cite{oommen2022learning,liu2022causality}. The input to the branch network remains unchanged from the earlier discussion. It is important to note that during the prediction stage, the initial data of $\eta(t)$ containing the first $n_h$ history terms is provided. The predictions generated by the network itself are concatenated and passed to the trunk network, along with the time step, for making future five predictions. 
\bigbreak
\noindent
\textbf{Self adaptive DeepONet}\\
In the optimization process, certain query points need to be penalized more than others to meet specific constraints such as initial conditions or boundary conditions. In such scenarios, the use of non-uniform training point weights, designed appropriately, can enhance accuracy. These penalizing parameters can be manually adjusted, but this approach is often time-consuming or requires adaptive decision-making during DeepONet training~\cite{mcclenny2020self,kontolati2022influence}. Alternatively, these parameters in the loss function can be updated alongside the network parameters using gradient descent. The adjusted loss function is defined as:
\begin{align}
    \mathcal L(\bm \theta, \bm \lambda) = \frac{1}{N}\sum_{i = 1}^{N} g(\bm\lambda)|u_i (\xi)- \mathcal G_{\mathbf\theta}(\mathbf{v}_i)(\xi)|^2,
\end{align}
where $g(\lambda)$ is non-negative, strictly increasing self-adaptive mask function, and $\bm \lambda = \{\lambda_1, \lambda_2, \cdots \lambda_j\}$ are $j$ self-adaptive parameters, each associated with an evaluation point, $\xi_j$. In this study, we set $j$ equal to the number of time points, denoted as $n_t$. These parameters are subjected to the constraint of monotonically increasing values and are always positive. Usually, in a neural network, we aim to minimize the loss function concerning the network parameters, denoted as $\boldsymbol{\theta}$. However, in this approach, we go a step further and also maximize the loss function with respect to the trainable hyperparameters using a gradient descent/ascent method. The adjusted objective function is defined as:
\begin{align}
    \min_{\bm \theta} \max_{\bm \lambda}\mathcal L(\bm \theta, \bm \lambda).
\end{align}
The self-adaptive weights are updated using the gradient descent method, such that
\begin{align}\label{eq:def_lambda}
    \bm \lambda^{k+1} = \lambda^{k} + \eta_{\lambda}\nabla_{\bm \lambda}\mathcal L(\bm \theta, \bm \lambda),
\end{align}
where $\eta_{\bm \lambda}$ is the learning rate of the self-adaptive weights and
\begin{align}\label{eq:gradient_lambda}
    \nabla_{\lambda_i}\mathcal L = \left[g'(\lambda_i)(u_i (\xi)- \mathcal G_{\mathbf\theta}(\mathbf{v}_i)(\xi))^2\right]^{T}.
\end{align}
Therefore, if $g(\lambda_i)>0$, $\nabla_{\lambda_i}\mathcal L$ would be zero only if the term $(u_i (\xi)- \mathcal G_{\mathbf\theta}(\mathbf{v}_i)(\xi))$ is zero. The introduction of self-adaptive weights in \cite{kontolati2022influence} has greatly enhanced the accuracy of predicting discontinuities or non-smooth features in the solution.

\bigbreak
\noindent
\textbf{Wavelet DeepONet}\\
In this section, we present an extension of the original DeepONet called Wavelet DeepONet (W-DeepONet), which draws inspiration from the effectiveness of approximating wavelet components in a time series. The architecture of W-DeepONet consists of a branch and a trunk network. The inputs for the branch network and trunk networks are the concatenated wavelet coefficients (approximation and detail) of the wave elevation, denoted as $\bm{\eta}_i = [\bm{\eta}_{i,A}, \bm{\eta}_{i,D}]$, and the approximation coefficients of the time coordinates, denoted as $\bm y_A$, respectively. The proposed W-DeepONet involves three steps:

\noindent (1) We perform discrete wavelet transforms on $\bm{\eta}_i$, $\bm y$, and the response $\mathbb G(\bm \eta_{i})(\bm y)$, yielding the approximation wavelet coefficients $\bm{\eta}_{i,A}$, $\bm y_A$, $\mathbb G(\bm \eta_{i,A})$, and detail wavelet coefficients $\bm{\eta}_{i,D}$, $\bm y_D$, $\mathbb G(\bm \eta_{i,D})$.

\noindent (2) The branch network encodes $\bm{\eta}_{i} = [\bm{\eta}_{i,A}$ and $\bm{\eta}_{i,D}]$, while the trunk network encodes the information about the coordinates $\bm y_A$, where the solution operators are evaluated as follows: 
\begin{align} \label{eq:output_deeponets_A}
    \mathbb G_{\bm{\theta}}(\bm{\eta}_{i,A})  &= \sum_{k = 1}^{q/2}b_k(\eta_{i,A}(t_1), \eta_{i,D}(t_1), \ldots, \eta_{i,A}(t_{n_{sen}}), \eta_{i,D}(t_{n_{sen}}))\cdot t_k(\bm y_A).
\end{align} 
\begin{align} \label{eq:output_deeponets_D}
    \mathbb G_{\bm{\theta}}(\bm{\eta}_{i,D}) &= \sum_{k = \frac{q}{2} + 1}^{q}b_k(\eta_{i,A}(t_1), \eta_{i,D}(t_1), \ldots, \eta_{i,A}(t_{n_{sen}}), \eta_{i,D}(t_{n_{sen}}))\cdot t_k(\bm y_A).
\end{align} 
The expressions for the trunk network outputs are given by Eqs.~ \ref{eq:output_deeponets_A} and \ref{eq:output_deeponets_D}. The trainable parameters of W-DeepONet denoted as $\bm{\theta}$ in Eqs.~\ref{eq:output_deeponets_A} and \ref{eq:output_deeponets_D}, are obtained by minimizing the following loss function defined in Eq.~\ref{eq:L1_loss_W_deeponet} as:
\begin{align} \label{eq:L1_loss_W_deeponet}
    \mathcal L_2 &= \sum_{i =1}^n \sum_{j =1}^p \big(\mathbb G(\bm{\eta}_{i,A})(\bm y_{j,A}) - \mathbb G_{\bm{\theta}}(\bm{\eta}_{i,A})(\bm y_{j,A})\big)^2+ \sum_{i =1}^n \sum_{j =1}^p \big(\mathbb G(\bm{\eta}_{i,D})(\bm y_{j,A}) - \mathbb G_{\bm{\theta}}(\bm{\eta}_{i,D})(\bm y_{j,A})\big)^2.
\end{align}

\noindent (3) Once the solution operators $\mathbb G_{\bm{\theta}}(\bm{\eta}_{i,A})(\bm y_{j,A})$ and $\mathbb G_{\bm{\theta}}(\bm{\eta}_{i,D})(\bm y_{j,A})$ are obtained, the inverse wavelet transform is performed to obtain the predicted responses in the time domain.

\subsection{Fourier Neural Operator}

The Fourier neural operator (FNO), introduced by Li et al.~\cite{li2020fourier}, replaces the kernel integral operator with a convolution operator defined in Fourier space. This operator takes input functions defined on a regularly spaced lattice grid and produces the desired field on the same grid points. In FNO, the network parameters are defined and learned in Fourier space instead of physical space. In other words, the coefficients of the Fourier series of the output function are learned from the data. FNO can be seen as an extension of DeepONet, where a convolutional neural network is employed in the branch network to approximate the input functions and Fourier basis functions are used in the trunk network. The network of FNO consists of three main components:\\
(1) The input function $\mathbf{\eta}(t)$ is lifted to a higher-dimensional representation $\hb(x,0)$ using a lifting layer denoted as $\mathcal{P}$. This lifting layer is typically parameterized by a linear transformation or a shallow neural network.\\
(2) The neural network architecture is formulated iteratively as $\hb(x,0)\rightarrow \hb(x,1)\rightarrow\hb(x,2)\rightarrow \cdots \rightarrow \hb(x,L)$. Here, $\hb(x,j)$ for $j=0,\cdots, L$ represents the values of the architecture at each layer. Each layer is defined as a nonlinear operator by applying a sum of Fourier transformations and a bias function, as shown in Eq.~\ref{eq:FNO}. 
\begin{align}
\nonumber\hb(x,j+1)=&\mcL_j^{FNO}[\hb(x,j)]\\
:=&\sigma\left(W_j\hb(x,j)+\mathcal{F}^{-1}[R_j\cdot \mathcal{F}[\hb(\cdot,j)]](x)+ \cb_j\right).\label{eq:FNO}
\end{align}
The activation function is denoted as $\sigma$, and $W_j$, $\cb_j$, and $R_j$ are the trainable parameters specific to the $j$-th layer, allowing each layer to have different kernels, weights, and biases.\\
(3) Finally, the output $\mathbf{u}(x)$ is obtained by projecting $\hb(x,L)$ through a local transformation operator layer denoted as $\mathbf{Q}$.

\subsection{Wavelet Neural Operator}

The wavelet neural operator (WNO) introduced by Tripura et al.~\cite{tripura2022wavelet} adopts a learning approach in the wavelet space, which allows for localized frequency and spatial information. This enables the network to effectively capture patterns in images and signals. In contrast to the Fourier integrals used in FNO, WNO employs wavelet integrals to capture the spatial characteristics of a signal and handle complex boundary conditions. WNO has demonstrated its capability to handle domains with both regular and complex geometries. It has been successfully applied to learn solution operators for highly nonlinear partial differential equations (PDEs) that involve discontinuities and abrupt changes in the solution domain and boundary. 

Motivated by the idea of self-adaptive DeepONet (SA-DeepONet), in this work, we have introduced self-adaptive WNO. In principle, the idea of this extension is similar to SA-DeepONet, however, to implement it in practice, we have modified Eqs.~\ref{eq:def_lambda} and \ref{eq:gradient_lambda} such that:
\begin{align}
    \bm \lambda^{k+1} &= \lambda^{k} - \eta_{\lambda}\nabla_{\bm \lambda}\mathcal L(\bm \theta, \bm \lambda),\\
    \nabla_{\lambda_i}\mathcal L &= \left[\mathcal G_{\mathbf\theta}(\mathbf{v}_i)(\xi))^2 - g'(\lambda_i)(u_i (\xi)\right]^{T},
\end{align}
respectively.

\section{Numerical Results}
\label{sec:results}

\begin{figure}[H]
\centering
\includegraphics[width=4 in]{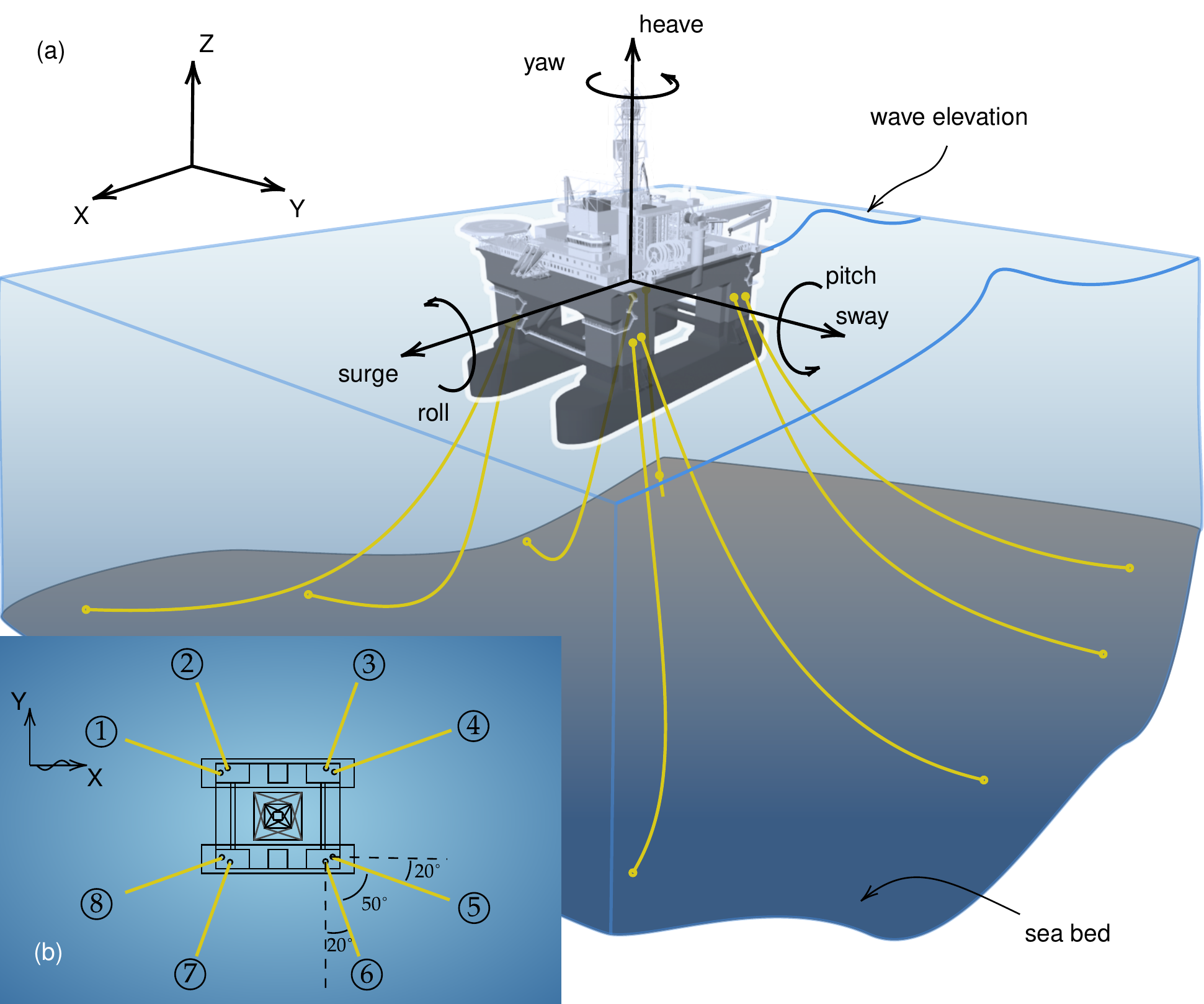}\\
\caption{A schematic representation of a semi-submersible platform with the mooring system: (a) side view, (b) top view}\label{semi-submersible}
\end{figure}

In this section, we have presented the performance of all the neural operators and their extensions discussed in Section~\ref{sec:methods} for a semi-submersible platform coupled with its mooring system for zero and non-zero initial conditions. Additionally, we have also compared the results of the neural operators with gated recurrent units (GRU). A schematic representation of the structure is shown in Fig.~\ref{semi-submersible}, which is composed of a floating offshore platform anchored by eight mooring lines distributed in four groups. The weight of the platform is $53.4\times10^3$ tons, the length is $105$ m, and the working water depth is $234$ m. The four groups of mooring lines are evenly distributed at an angle of $90^{\circ}$. For each group, the angle between two mooring lines is $50^{\circ}$. The properties of the mooring lines are listed in Table~\ref{mooring_parameter}. The platform is modeled as a rigid body with six degrees of freedom ($3$ translations and $3$ rotations). The platform does not deform under external loading. The dynamics of such rigid body are described by the laws of kinematics and Newton's second law. The platform motions in waves are defined by Response Amplitude Operators (RAOs). Orcaflex utilizes hydrodynamic data generated by WAMIT, which is based on the linear and second-order potential theory, and the velocity potential is computed by the boundary element method. However, it is important to note that this analysis does not account for significant viscous damping effects. To address nonlinear and breaking ocean waves, the incorporation of a viscous model is imperative~\cite{del2021learning}. A finite element model is used to simulate the mooring lines. Each mooring line is segmented into straight, massless segments with nodes at both ends. These segments represent the axial and torsional properties of the line, while other properties such as mass, weight, and buoyancy are all lumped to the nodes. The hydrodynamic loads on lines are calculated by Morison's equation with defined coefficients~\cite{manual2012online}. A time-domain simulation that uses an explicit Euler integration technique with a constant time step is chosen to solve the equation of motion.

\begin{table}[H]
\renewcommand\arraystretch{1.05}
\centering
   \caption{Properties of mooring system}
   \label{mooring_parameter}
   \scriptsize
   \begin{tabular}{ccccccccc}
   \Xhline{2\arrayrulewidth}
    Type  & Total length   & Diameter   & Mass per unit length& Axial stiffness  & Breaking strength  & Added mass   & Inertia  & Drag  \\
                &(m)   & (mm) & (kg$/$m) &(kN)& (kN)  &coefficient & coefficient&coefficient\\
   \Xhline{2\arrayrulewidth}
   Chain   & 1700  & 158.76  & 154.53& 712.7e3   & 7200 & 1   & 0.07 & 1  \\
    \Xhline{2\arrayrulewidth}
   \end{tabular}
\end{table}

\begin{figure}[t]
\centering
\includegraphics[trim=0cm 0.5cm 8cm 2cm,clip=true,width=3.8 in,angle=-90]{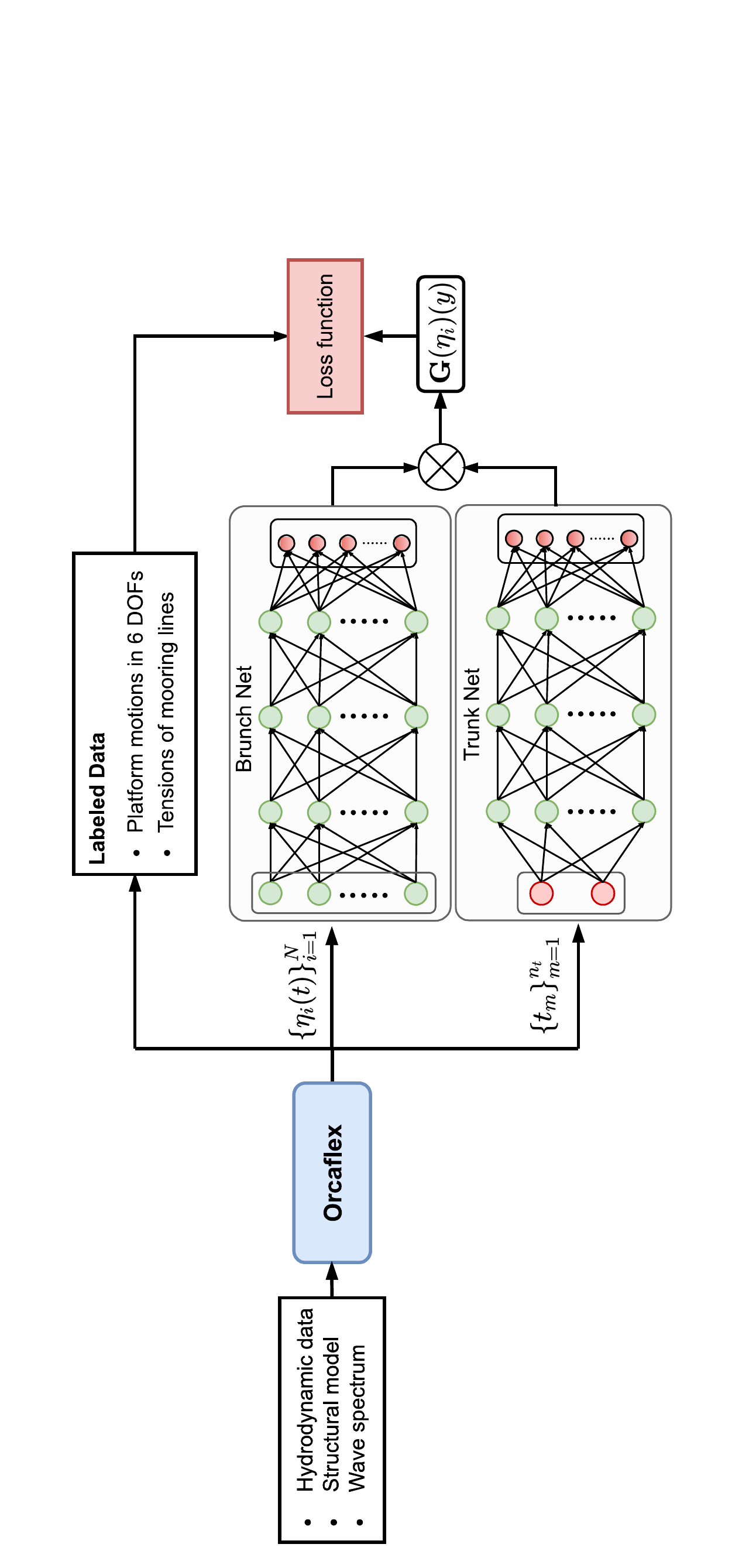}\\
\caption{The workflow of the setup is presented. The discretized JONSWAP spectrum, the hydrodynamic data (e.g., the added mass, potential damping and wave load RAO), and the model of the floating structure is provided as input to the Orcaflex, which outputs the wave elevations, $\eta(t)$, the six degrees of freedom (heave, pitch, surge, roll, yaw, and sway), and the tensions of mooring lines on the temporal scale. The outputs of Orcaflex are used to train the neural operators as shown in Fig.~\ref{Figure_neural_operator}.}\label{workflow1}
\end{figure}

The platform is assumed to be subjected to an irregular wave characterized by the JONSWAP spectrum. The realizations of the wave elevations under different sea states are constructed by the superposition technique of linear wave components~\cite{hu2018laplace}. The frequency of each wave component is obtained from the discretized JONSWAP spectrum. The heading angle is always considered to be $0^{\circ}$. In this wave approach condition, the surge, heave and pitch are the most dominate motions, which imply that only the surge, heave, and pitch of the platform should be investigated. Therefore, only these three DOFs are the focus of the current effort. Due to symmetry and similar characteristics of mooring lines, only the top tension of mooring line \#1 is learned. The labeled datasets are generated by running a commercial software \textit{Orcaflex}, which carries out the dynamic coupling analysis of the semi-submersible platform and the mooring systems. The workflow for the setup is shown in Fig.~\ref{workflow1}. We have not considered experimental or field data in the present work. If newly measured environmental and response data from in-service platforms are available, the accuracy of the surrogate model can be continuously improved and updated. DeepONet and its extensions and GRU have been implemented using $\texttt{Tensorflow}$ \cite{agarwal2016ten}, while FNO and WNO along with their extensions have been implemented using the $\texttt{PyTorch}$ \cite{paszke2019pytorch}. Moreover, we have employed the Adam optimizer and random norm initialization, unless otherwise mentioned. The experiments carried out in this work are listed in Table~\ref{table:cases}. Details about the network architecture adopted in each case for every operator are presented in Supplementary Section $3$. For a visual representation, additional plots of the predictions compared against the ground truth are shown in Supplementary Section $1$. The error metrics used to measure the accuracy of the models can be expressed as:
\begin{align}
\begin{split}
\text{MSE} = \frac{1}{n} \sum_{i=1}^{n} ( y_i - \hat{y_i} )^2\;\; \text{and}\;\;  \text{Relative}\;\; \mathcal L_2 = \frac{||y_i - \hat{y_i}||_2}{||y_i||_2} \;\;\;\; \text{for}\;\; i =(1,n),
\end{split}
\end{align}
where $n$ is the number of samples, $y_i$ is the true value of the $i^{th}$ sample, and $\hat{y_i}$ is the predicted value of the $i^{th}$ sample.

\begin{figure}[H]
\centering
\includegraphics[width=6 in]{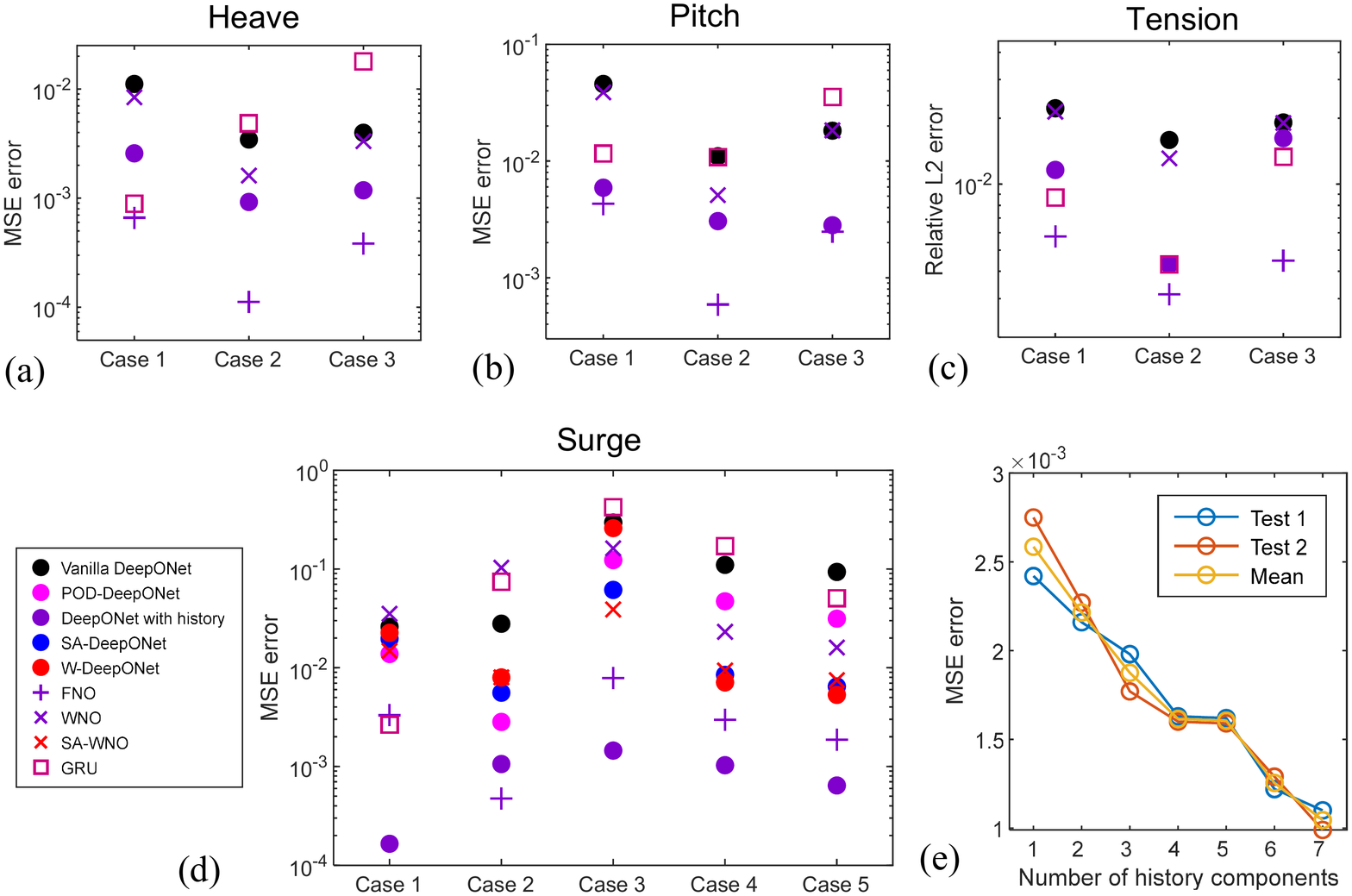}\\
\caption{Error in the test cases for (a) heave and (b) pitch motions, (c) tension lines, and (d) surge motion for all the neural operators and their extensions for cases $1$--$5$, that considers zero initial conditions. (e) Influence of the number of history terms, $n_h$ in the test accuracy for case $5$ to approximate the surge motion with history DeepONet that employs historical states as basis enhancements along with the temporal coordinates in the trunk net. Test$1$ and Test$2$ represent two independent trials. In all the plots, (a)-(d), $n_h = 5$.} \label{fig:mse_error_surge_zeroInit}
\end{figure}

\begin{figure}[H]
\centering
\includegraphics[width=6 in]{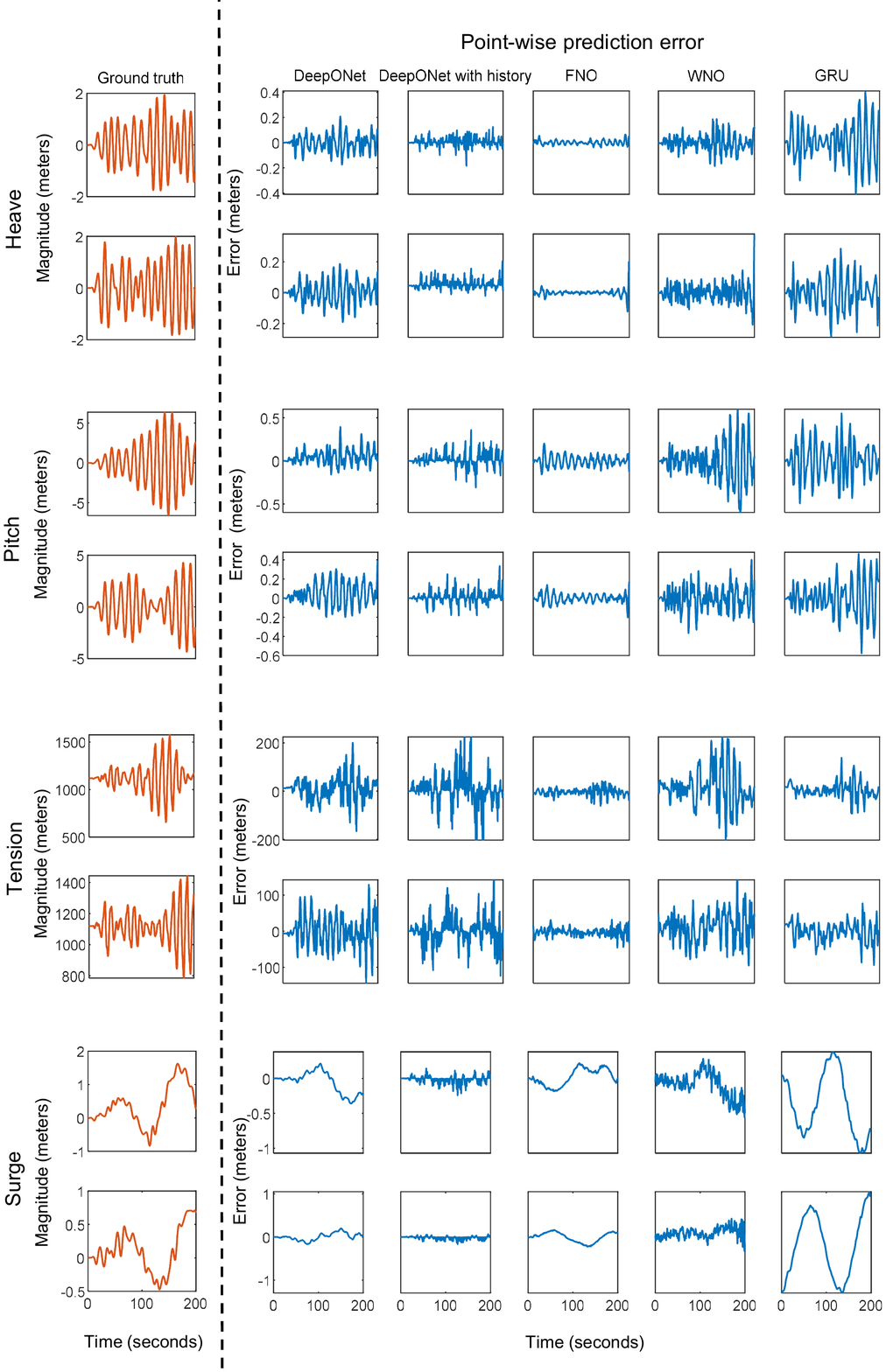}\\
\caption{Pointwise error plots of heave and pitch motions, tension lines, and surge motion for two representative test samples drawn from case $3$ that considers zero initial conditions. The ground truth is shown in the left column and the pointwise error for vanilla-DeepONet, DeepONet with $5$ historical states, FNO, vanilla WNO, and GRU are presented in the right section.}\label{fig:err_zeroInit}
\end{figure}

For cases $1$ and $6$, we have considered the sea state with significant wave height, $H_s=5.5$m, and zero-crossing wave period, $T_z=13.5$s. For cases $2$ and $7$, we have generated samples for the sea states with combinations of $H_s$ and $T_z$ drawn from the sets: $H_s = [3.5, 4.5]$ and $T_z = [7.5, 8.5, 9.5, 10.5]$, of which the dataset with $H_s = 3.5$m and $T_z = 8.5$s is considered for validation and testing and the rest is used for training the network. For cases, $3$ and $8$, the number of samples generated for each combination of $H_s$ and $T_z$ are shown in Table~\ref{tab:all_sea_states}. From the generated samples, the data for $H_s = 2.5$m and $T_z = 9.5$s is considered for testing and validation, while the remaining combinations are used for training the operator networks. By leveraging $40$ parallel threads, the Orcaflex software efficiently simulated $1,000$ samples in approximately half an hour, employing the mechanical models and computational methods described earlier. For instance, in Case $1$, we generated a total of $8,000$ samples, taking approximately $4$ hours to complete the data generation process for training and testing. The errors of the test cases obtained for each case for all the discussed approaches are shown in Fig.~\ref{fig:mse_error_surge_zeroInit}. Among all the vanilla forms of the neural operators, FNO demonstrates an overall high accuracy in response prediction for all cases. The low prediction accuracy of surge motions employing DeepONet and WNO may be attributed to its broad-band spectrum feature. This conclusion is consistent with that given by Ye et al.~\cite{ye2022emd}, which states that the spectrum bandwidth deteriorates the prediction accuracy of DNNs. To that end, we increased the number of samples generated for all combinations of $H_s$ and $T_z$ for surge motion in cases $4$ and $5$. As expected, the generalization of the neural operators improves in cases $4$ and $5$. Furthermore, we also observe a marked improvement in the accuracy of DeepONet when $5$ history terms are recursively updated in the trunk net to improve the computation of the basis function which is obtained as outputs from the trunk net. In Fig. \ref{fig:mse_error_surge_zeroInit}(a)-(d), we show the error (MSE for heave, pitch, and surge, and relative $L_2$ for tension) between the studied models and ground truth. We observe that all the operators with their vanilla architectures have similar accuracy for heave, pitch and tension. However, for surge, the vanilla architectures fail to the capture the response, hence we investigate it further with modified architecture, and conclude that DeepONet with $5$ historical states outperform all other operators. Motivated by the accuracy of DeepONet with historical states, we conducted additional experiments by varying the number of historical terms, $n_h$ in the trunk net. Additionally, in \ref{fig:mse_error_surge_zeroInit}(e) we present the improvement in the prediction accuracy with varying $n_h$ for approximating the surge motion of case $3$ for two independent trials. Error plots of heave, pitch, tension lines, and surge for two representative test samples with zero initial conditions for case $3$ are shown in Fig.~\ref{fig:err_zeroInit} for the best extension of each operator. Furthermore, we also observe that W-DeepONet and SA-WNO are significantly more accurate than their vanilla counterparts for the surge motion. 

\begin{table}[H]\label{tab:cases}
\renewcommand\arraystretch{1.15}
\centering
  \caption{Description of the datasets considered for the experiments performed. The time length considered is $200$s with $\Delta t=0.2$s, hence $n_t = 1000$ temporal points per signal. For each experiment, $N_{tr}$, $N_{val}$, and $N_{te}$ denote the number of training, validation, and testing samples, respectively.}\label{table:cases}
  \begin{tabular}{ccl}
  \Xhline{2\arrayrulewidth}
     & Case \# & Description \\
  \Xhline{2\arrayrulewidth}
  \multicolumn{1}{l|}{\multirow{10}{*}{\begin{turn}{90}Zero initial conditions\end{turn}}} & \multicolumn{1}{c|}{Case 1} & Training and testing on same sea state.\\
  \multicolumn{1}{l|}{} & \multicolumn{1}{l|}{}      & Heave/Pitch: $N_{tr} = 800$, $N_{val} = 100$, $N_{te} = 100$\\
    \multicolumn{1}{l|}{} & \multicolumn{1}{l|}{}      & Tension: $N_{tr} = 2800$, $N_{val} = 100$, $N_{te} = 100$\\
    \multicolumn{1}{l|}{} & \multicolumn{1}{l|}{}      & Surge: $N_{tr} = 7200$, $N_{val} = 400$, $N_{te} = 400$\\
 \multicolumn{1}{l|}{} & \multicolumn{1}{c|}{Case 2} & Training on $7$ sea states \& testing on an unknown sea state. \\
 \multicolumn{1}{l|}{} &    \multicolumn{1}{l|}{}      &$N_{tr} = 14\small{,}000$, $N_{val} = 1\small{,}000$, $N_{te} = 1\small{,}000$\\
 \multicolumn{1}{l|}{} & \multicolumn{1}{c|}{Case 3} & Training on $72$ sea states with limited samples \& testing on an unknown sea state.\\
 \multicolumn{1}{l|}{} &   \multicolumn{1}{l|}{}       &$N_{tr} = 959$, $N_{val} = 23$, $N_{te} = 20$\\
 \multicolumn{1}{l|}{} &  \multicolumn{1}{c|}{Case 4} & Training on $72$ sea states \& testing on an unknown sea state. \\
 \multicolumn{1}{l|}{} &    \multicolumn{1}{l|}{}      &$N_{tr} = 14\small{,}400$, $N_{val} = 100$, $N_{te} = 100$\\
 \multicolumn{1}{l|}{} &  \multicolumn{1}{c|}{Case 5} & Training on $72$ sea states \& testing on an unknown sea state. \\
 \multicolumn{1}{l|}{} &   \multicolumn{1}{l|}{}       &$N_{tr} = 21\small{,}600$, $N_{val} = 100$, $N_{te} =200$\\
  \Xhline{2\arrayrulewidth}
  \multicolumn{1}{l|}{\multirow{10}{*}{\begin{turn}{90}Varying initial conditions\end{turn}}} & \multicolumn{1}{c|}{Case 6} & Training and testing on same sea state.\\
  \multicolumn{1}{l|}{} &    \multicolumn{1}{l|}{}      &Heave/Pitch: $N_{tr} = 800$, $N_{val}$ = $100$, $N_{te} = 100$\\
      \multicolumn{1}{l|}{} & \multicolumn{1}{l|}{}      & Tension: $N_{tr} = 2800$, $N_{val} = 100$, $N_{te} = 100$\\
    \multicolumn{1}{l|}{} & \multicolumn{1}{l|}{}      & Surge: $N_{tr} = 7200$, $N_{val} = 400$, $N_{te} = 400$\\
 \multicolumn{1}{l|}{} & \multicolumn{1}{c|}{Case 7} & Training on $7$ sea states \& testing on an unknown sea state. \\
 \multicolumn{1}{l|}{} &    \multicolumn{1}{l|}{}      &$N_{tr} = 14\small{,}000$, $N_{val} = 1\small{,}000$, $N_{te} = 1\small{,}000$\\
 \multicolumn{1}{l|}{} & \multicolumn{1}{c|}{Case 8} & Training on $72$ sea states with limited samples \& testing on an unknown sea state.\\
 \multicolumn{1}{l|}{} &   \multicolumn{1}{l|}{}       &$N_{tr}$ = $959$, $N_{val}$ = $23$, $N_{te} =20$\\
 \multicolumn{1}{l|}{} &  \multicolumn{1}{c|}{Case 9} & Training on $72$ sea states \& testing on an unknown sea state. \\
 \multicolumn{1}{l|}{} &    \multicolumn{1}{l|}{}      &$N_{tr}$ = $14\small{,}400$, $N_{val}$ = $100$, $N_{te} =100$\\
 \multicolumn{1}{l|}{} &  \multicolumn{1}{c|}{Case 10} & Training on $72$ sea states \& testing on an unknown sea state. \\
 \multicolumn{1}{l|}{} &   \multicolumn{1}{l|}{}       &$N_{tr}$ = $21\small{,}600$, $N_{val}$ = $100$, $N_{te} =200$\\
  \Xhline{2\arrayrulewidth}
  \end{tabular}
\end{table}

\begin{table}[htbp!]
\renewcommand\arraystretch{1.25}
\scriptsize
\centering
   \caption{The sea states described in Table~\ref{table:cases} are defined by their wave height ($H_s$) and zero crossing wave period ($T_z$). For cases 1 and 6, the blue-marked case is considered, while the green-marked zone is used for cases 2 and 7. The table includes all relevant sea states for cases 3-5 and 8-10. In cases 3 and 8, which test the model's performance with sparse data, the number of samples generated for each sea state is equal to the value of the cell at the intersection of the components, $H_s$ and $T_z$. These sea states are specific to the North Atlantic Ocean.}
   \label{tab:all_sea_states}
   \begin{tabular}{cccccccccccc}
   \Xhline{2\arrayrulewidth}
    \multirow{2}{*}{$H_s$ (m)}  & \multicolumn{11}{ c }{$T_z$ (s)} \\
     \cline{2-12}
              & 4.5 & 5.5 &6.5 & \cellcolor{green!40}7.5 & \cellcolor{green!40}8.5 &\cellcolor{green!40}9.5 &\cellcolor{green!40}10.5 &11.5 &12.5 &\cellcolor[HTML]{56ADE8}13.5 & Total \\
   \Xhline{2\arrayrulewidth}
    0.5        & 3 & 14 &21 &13 &4 &1 & & & & & 56 \\
    1.5        &   & 13 &54 &75 &47 &16 &4 &1 & & & 210 \\
    2.5        &   & 5 &33 &79 &80 &43 &14 &4 & 1& & 259 \\
    \cellcolor{green!40}3.5        &   & 1 &14 & 46 & 65 & 48 & 21 &6 & 2& & 203 \\
    \cellcolor{green!40}4.5        &   &  &5 &21 & 38 & 34 & 19 &7 & 2& & 126 \\
    \cellcolor[HTML]{56ADE8}5.5        &   &  &2 &9 &19 &20 &13 & 5&2 & 1 &71 \\
    6.5        &   &  &1 &3 &8 &11 &8 & 4&2 &1 &38 \\
    7.5        &   &  &  &1 &4 &5  &4 & 2&1 &  &17 \\
    8.5        &   &  &  &1 &2 &3  &3 & 2&1 &  &12 \\
    9.5        &   &  &  &  &1 &1  &2 & 1&1 &  &6 \\
   10.5        &   &  &  &  &  &1  &1 & 1&  &  &3 \\
   11.5        &   &  &  &  &  &   &1 &  &  &  &1 \\
   Total       &3  &33 &130 &248 &268  &183 &90 &33 & 12 &2  &1002 \\
   \Xhline{2\arrayrulewidth}
   \end{tabular}
\end{table}

\begin{figure}[t]
\centering
\includegraphics[width=6 in]{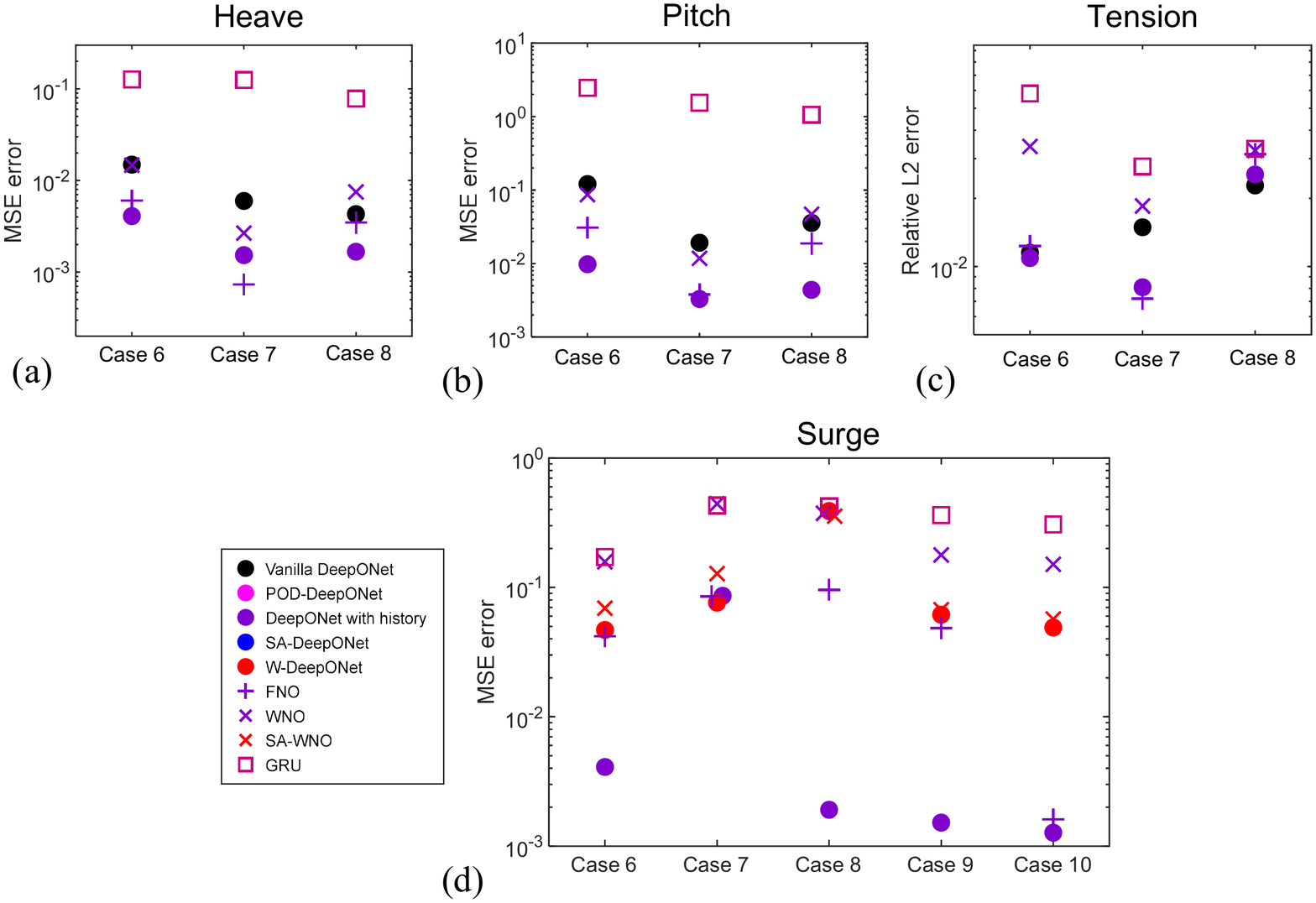}\\
\caption{Error in the test cases for (a) heave and (b) pitch motions, (c) tension lines, and (d) surge motion for all the neural operators and their extensions for cases $6$--$10$, that considers $6$ varying initial conditions. In all plots, $n_h = 5$ historical terms have been considered.}\label{fig:allApp_diffInit}
\end{figure}

\begin{figure}
\centering
\includegraphics[width=6 in]{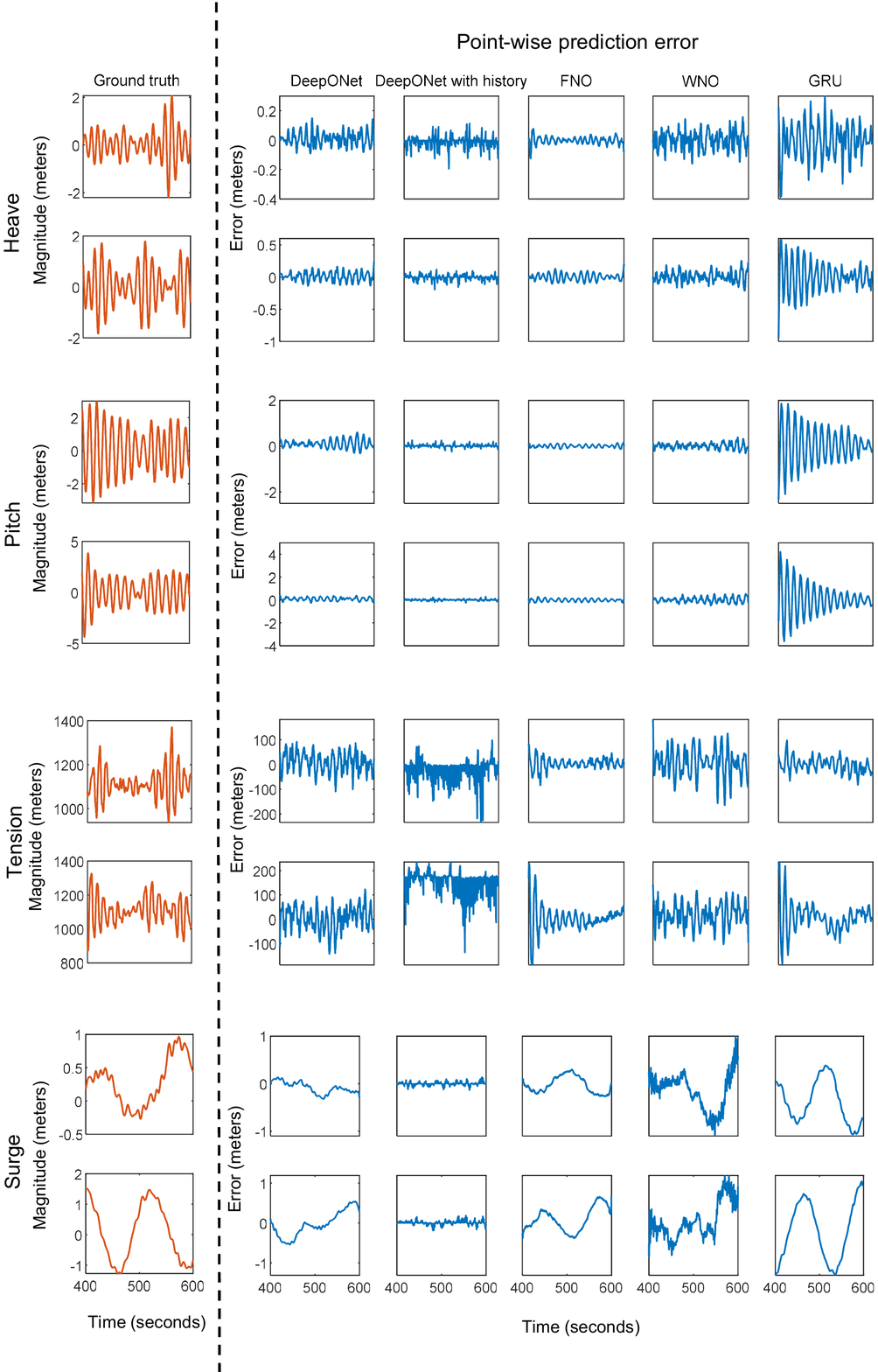}\\
\caption{Pointwise error plots of heave and pitch motions, tension lines, and surge motion for two representative test samples drawn from case $8$ that considers varying initial conditions. The ground truth is shown in the left column and the pointwise error for vanilla-DeepONet, DeepONet with $5$ historical states, FNO, vanilla WNO, and GRU are presented in the right section.}\label{fig:err_diffInit}
\end{figure}

\begin{figure}[t]
\centering
\includegraphics[width=6 in]{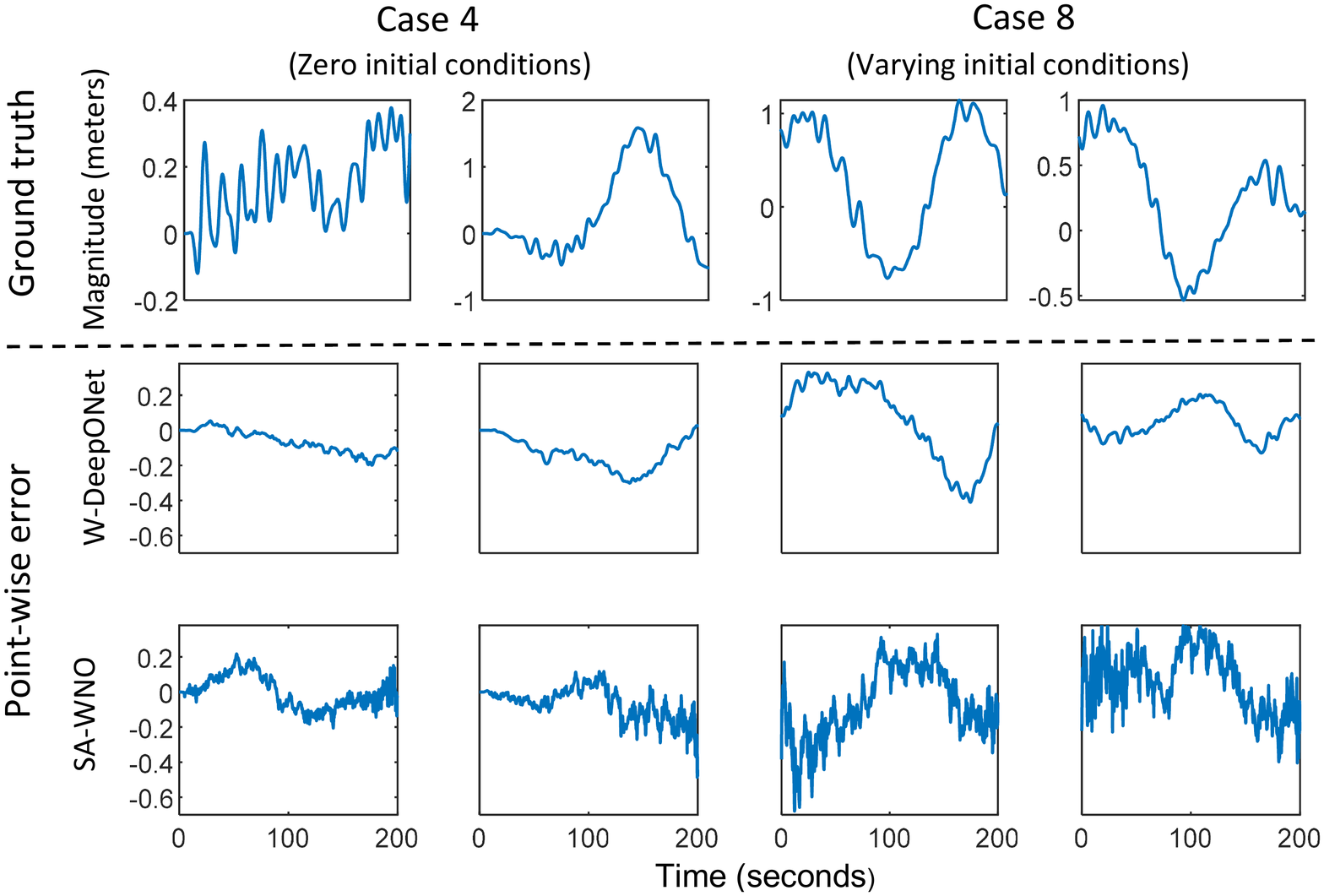}\\
\caption{Point-wise error for representative test samples with W-DeepONet and SA-WNO. The first two columns are for the solution with zero initial conditions and the last two columns are with varying initial conditions.}\label{fig:additionalplots}
\end{figure}

When a semi-submersible platform floats in the sea, a very likely situation is that the floater is not static. This situation will lead to non-zero initial conditions. On the other hand, if an ergodic signal is divided into multiple samples to learn the response, the initial conditions for these samples are also different. Thus, it is meaningful to establish a model to learn the operator which maps the wave elevation $\eta(t)$ and initial conditions $(\mathbf{x}(t_0),\dot{\mathbf{x}}(t_0))$ to the structural response $\mathbf{x}(t)$, as shown in Eqs.~\ref{eq:operator_platform} and \ref{eq:operator_mooring}. The experiments carried out for zero initial conditions of the responses (cases $1$--$5$) are repeated with different initial conditions (cases $6$--$10$). The architecture of the operators is modified such that:\\
1) Deeponet and its extensions have an additional branch network to input the initial condition. The additional branch net inputs the value of $6$ initial conditions ($3$ for displacement and $3$ for velocity), therefore $n_{sen}=6$ for the added branch network.\\
2) FNO, WNO, and its extensions consider the initial conditions as $6$ additional input functions ($3$ for displacement and $3$ for velocity). To that end, the initial conditions are repeated $n_t$ times, as the input functions have to be defined for every grid point.\\
\noindent The errors obtained in $N_{te}$ test cases for all approaches for cases $6$--$10$ are shown in Fig.~\ref{fig:allApp_diffInit}. We observe that all the neural operators are significantly more accurate than GRU. Considering the heave and pitch motion and tension lines, the vanilla neural operators have similar accuracy in the prediction of unseen test cases. However, for the surge motion, DeepONet with $5$ historical states and the time in the trunk net outperforms all the other operators and their extensions. We also observe a significant deterioration in the generalization accuracy of FNO for the surge motion with different initial conditions compared to the predictions with zero initial conditions (see~Fig.\ref{fig:mse_error_surge_zeroInit}). It can be concluded that FNO does not work well for capturing transient responses. Error plots of heave, pitch, tension lines, and surge for two representative test samples with different initial conditions for case $8$ are shown in Fig.~\ref{fig:err_diffInit}. Additionally, the error plots for two test samples with W-DeepONet and SA-WNO are shown in Fig.~\ref{fig:additionalplots}

Considering that capturing the surge motion is the most challenging, to evaluate the performance of the model, we compute the mean square error of predictions, and we report the mean and standard deviation of this metric based on three independent training trials in Supplementary Section 2.

\section{Summary}
\label{sec:summary_and_discussions}

In this work, we performed an extensive study for evaluating the accuracy and performance of three neural operators, namely the deep operator network (DeepONet), the Fourier neural operator (FNO), and the Wavelet neural operator (WNO) along with their extensions to approximate the three significant degrees of motion out of a total of six degrees for a semi-submersible platform under incident waves. Alongside neural operators, the study also considers the accuracy and performance of Gated recurrent units (GRU), which is a type of recurrent neural network that has been shown to learn long sequences more accurately than traditional neural networks. We have systematically studied the capability of the models for zero and non-zero initial conditions of displacement and velocities of the waves, for scenarios where the model was trained and tested with the data drawn for the same wave height and the time period, and also for situations where the model was tested with an unseen description of the wave height and time period. Additionally, we have proposed two novel extensions of the neural operators, namely the wavelet-DeepONet (W-DeepONet) and the self-adaptive WNO (SA-WNO). 

Of all the experiments studied in this work, cases $3$ and $8$ are the most challenging considering the availability of extremely limited labeled training data spanned over a broad range of wave height and time period. The training of the neural operators is followed by online inference, which could be used for real-time response prediction for any sea state and varying initial conditions of displacement and velocity. Our observations can be summarized as follows:
\begin{enumerate}
    \item For a small bandwidth of the frequency spectrum (heave, pitch, and top tensions of the platform's mooring line), and non-transient responses, FNO outperforms all the operators and their extensions (see Fig.~\ref{fig:mse_error_surge_zeroInit}(a)--(c)). FNO requires significantly less training time compared to the other neural operators considered in the study and also GRU. (see Table 9 in the supplementary).
    \item Considering transient responses in small bandwidth (different initial conditions for displacement and velocities), the prediction accuracy of FNO is drastically reduced compared to capturing non-transient responses. It is observed that in such scenarios DeepONet with $n_h=5$ historical terms achieves better accuracy (see Fig.\ref{fig:allApp_diffInit}(a)--(c)). 
    \item Addition of historical states to the trunk net along with the temporal co-ordinates for obtaining better basis functions results in a drastic improvement in the accuracy for approximating the surge motion with DeepONet (DeepONet with history, see Fig.~\ref{fig:mse_error_surge_zeroInit}(d) and \ref{fig:allApp_diffInit}(d)). The accuracy can be further improved by increasing the number of historical states (see Fig.~\ref{fig:mse_error_surge_zeroInit}(e)).
    \item We compare three neural operators (DeepONet, FNO, and WNO) with the recurrent neural network of the GRU type. The results show that all the neural operators are significantly more accurate than GRU except Case $1$. This indicates that traditional network architectures cannot be used for the efficient generalization of solutions across multiple differential equations. However, the neural operators generalize well across multiple differential equations. Additionally, GRUs have deteriorated performance when multiple functions are considered to be approximated at the same time (non-zero initial condition, see Fig.\ref{fig:allApp_diffInit}(a)--(c)).
    \item Since the input and output dimensions in FNO and WNO needs to be the same, enhancing the performance by adding historical information cannot be performed. 
    \item The neural operator extensions proposed through this study (W-DeepONet and SA-WNO) perform considerably better than their vanilla counterparts. The predictive accuracy of W-DeepONet is similar to SA-WNO in capturing both transient and non-transient responses for the surge motion. However, the prediction accuracy of SA-WNO is very sensitive to parameters, especially batch size.
    \item The improvement in the predictive accuracy and generalization of test cases with larger training data (case $5>$case $4>$case $3$ and case $10>$case $9>$case $8$) is established. 
    \item For 1000 samples, the trained neural operators predicted the outcomes in less than 10 seconds. By using 40 parallel threads, the Orcaflex software finished in roughly a half-hour. The findings demonstrated that a complete dynamic analysis was outperformed by neural operators with high precision in terms of the speed at which structural responses were delivered by more than two orders of magnitude.
    \item Additionally, depending on newly measured environmental and response data of in-service platforms, the neural operators based on the numerical data can be partially re-trained and continuously updated.
    \item Only severe sea states are studied in this paper. We anticipate that our preliminary study can accelerate the study of motion calculation of the offshore platform under more complex situations, such as extreme sea states and viscous numerical methods that consider nonlinear and breaking ocean waves.
\end{enumerate}

\subsection*{\textbf{Acknowledgement}}

QS would like to acknowledge the scholarship from the Dalian University of Technology for visiting Brown University, U.S.A. SG and GEK would like to acknowledge support by the DOE SEA-CROGS project (DE-SC0023191) and the OSD/AFOSR MURI grant FA9550-20-1-0358. The authors would like to acknowledge the computing support provided by the computational resources and services at the Center for Computation and Visualization (CCV), Brown University where all experiments were carried out. 

\subsection*{\textbf{Code availability}}
The codes for the two operator extensions (W-DeepONet and SA-WNO) developed as a part of this work are available on https://github.com/qianyingcao/Deep-neural-operators-for-the-response-of-floating-offshore-structures.

\bibliographystyle{elsarticle-num}
\bibliography{refer}

\newpage




\renewcommand{\thesection}{\small{S}\arabic{section}}
\renewcommand{\thefigure}{\small{S}\arabic{figure}}
\renewcommand{\thetable}{\small{S}\arabic{table}}
\setcounter{figure}{0}
\setcounter{table}{0}
\setcounter{section}{0}
\setcounter{page}{1}

\section{Representative plots of the response from neural operators.}\label{app:additional_plots}
\begin{figure}[H]
\centering
\begin{minipage}[t]{0.5\linewidth}
\centering
\includegraphics[width=3in]{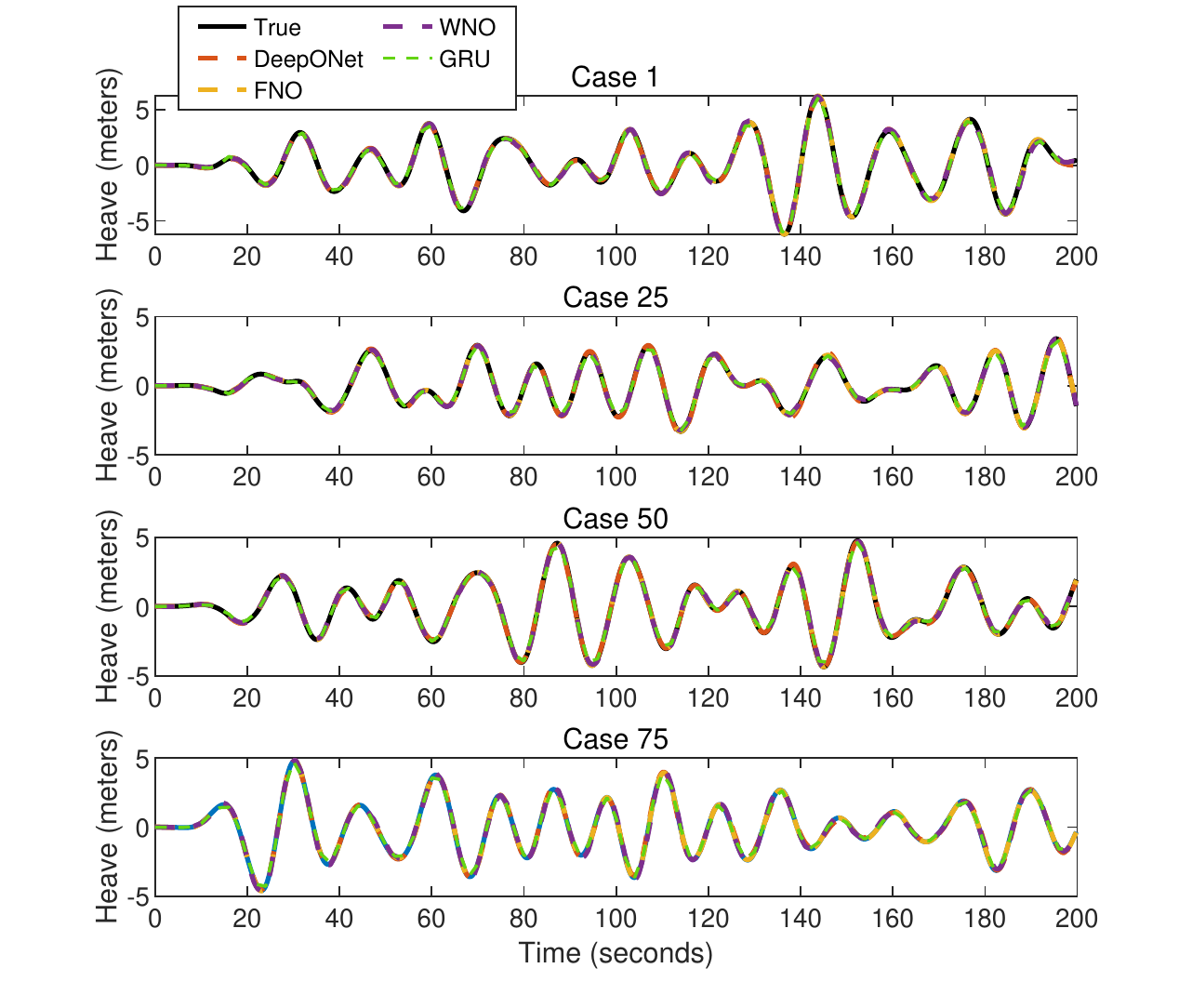}
\end{minipage}
\begin{minipage}[t]{0.5\linewidth}
\centering
\includegraphics[width=3in]{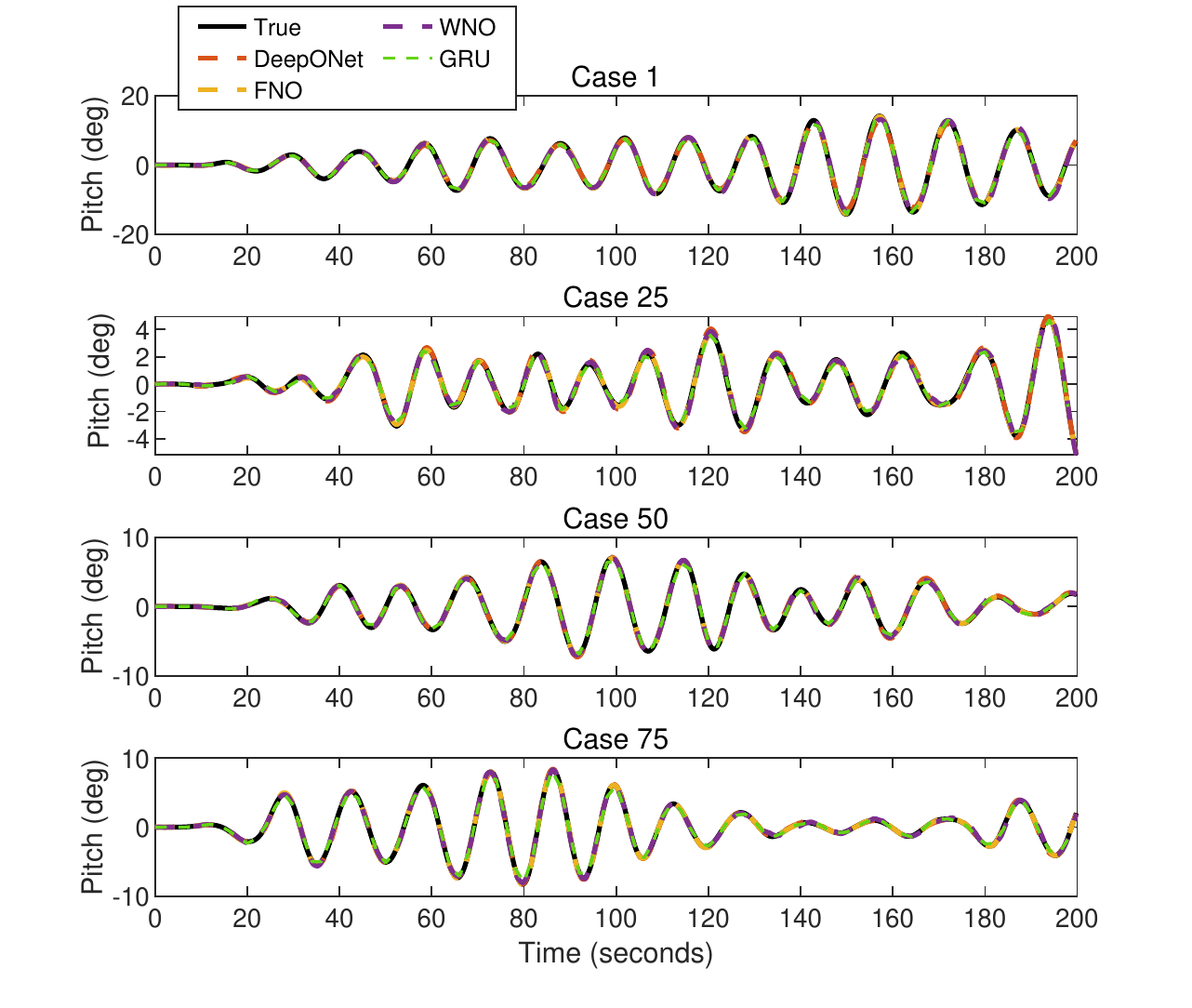}
\end{minipage}
\centering
\caption{Comparison of heave and pitch responses predicted by the four models, DeepONet, FNO, WNO, and GRU against the ground truth in Case 1. The plots on the left show the heave motion, while the plots on the right show the pitch motion.}\label{1state}
\end{figure}

\begin{figure}[H]
\centering
\begin{minipage}[t]{0.5\linewidth}
\centering
\includegraphics[width=3in]{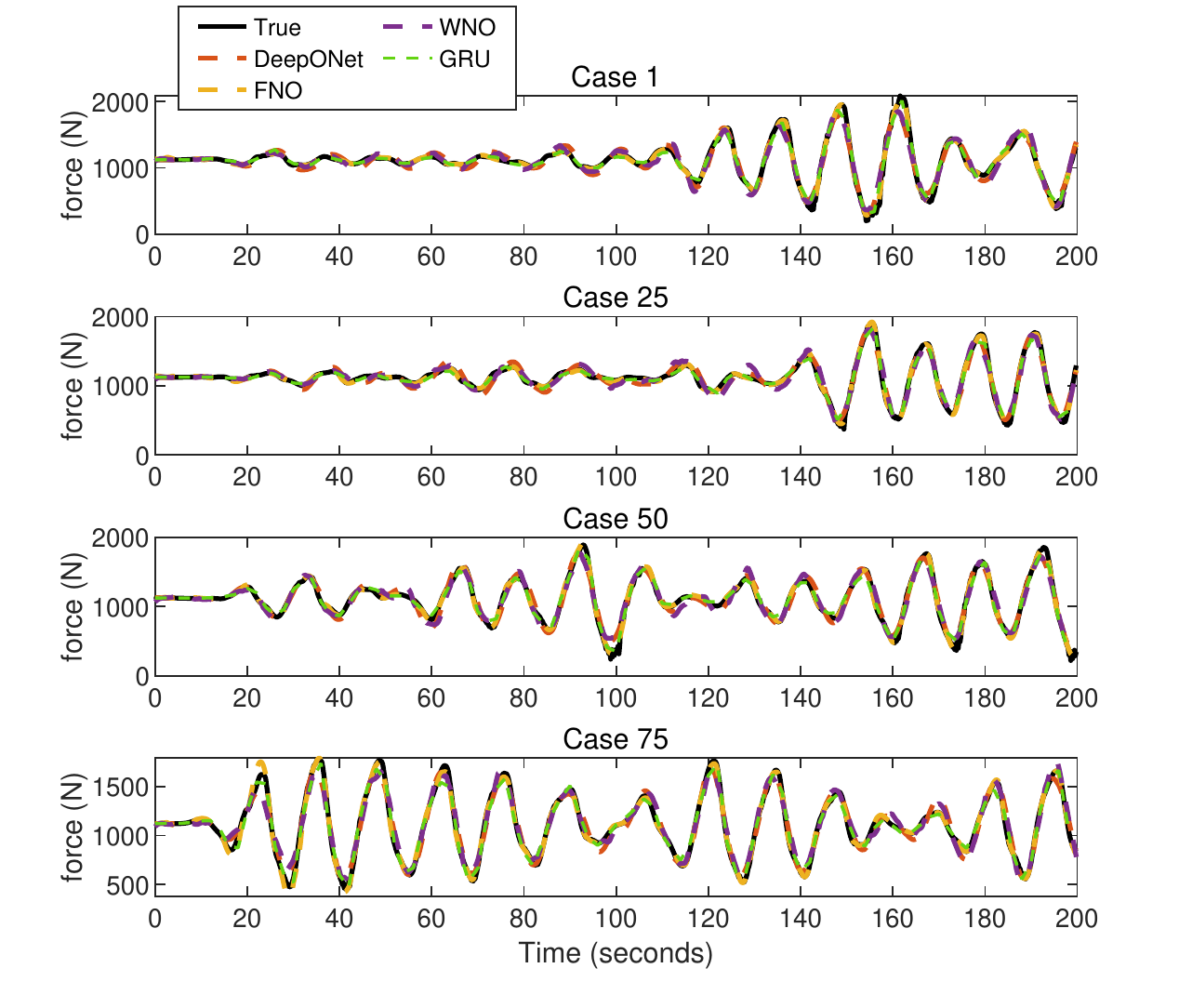}
\end{minipage}
\begin{minipage}[t]{0.5\linewidth}
\centering
\includegraphics[width=3in]{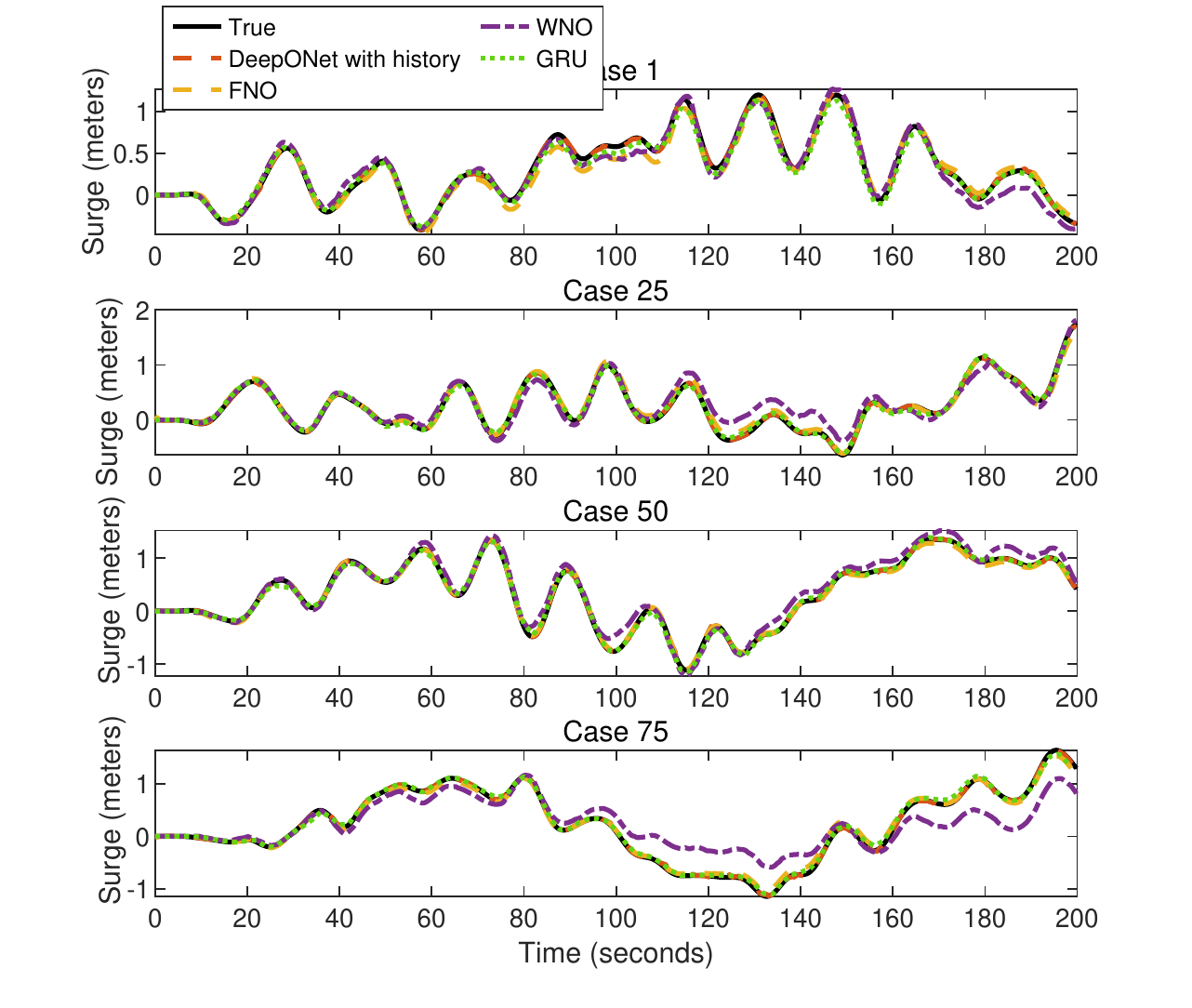}
\end{minipage}
\centering
\caption{Comparison of the top tension of mooring \#1 and surge response predictions obtained by the four models against the ground truth in Case 1. The plots on the left show the top tension, while the plots on right show the surge motion.}\label{moor_1state}
\end{figure}

\begin{figure}[H]
\scriptsize
\centering
\begin{minipage}[t]{0.5\linewidth}
\centering
\includegraphics[width=3in]{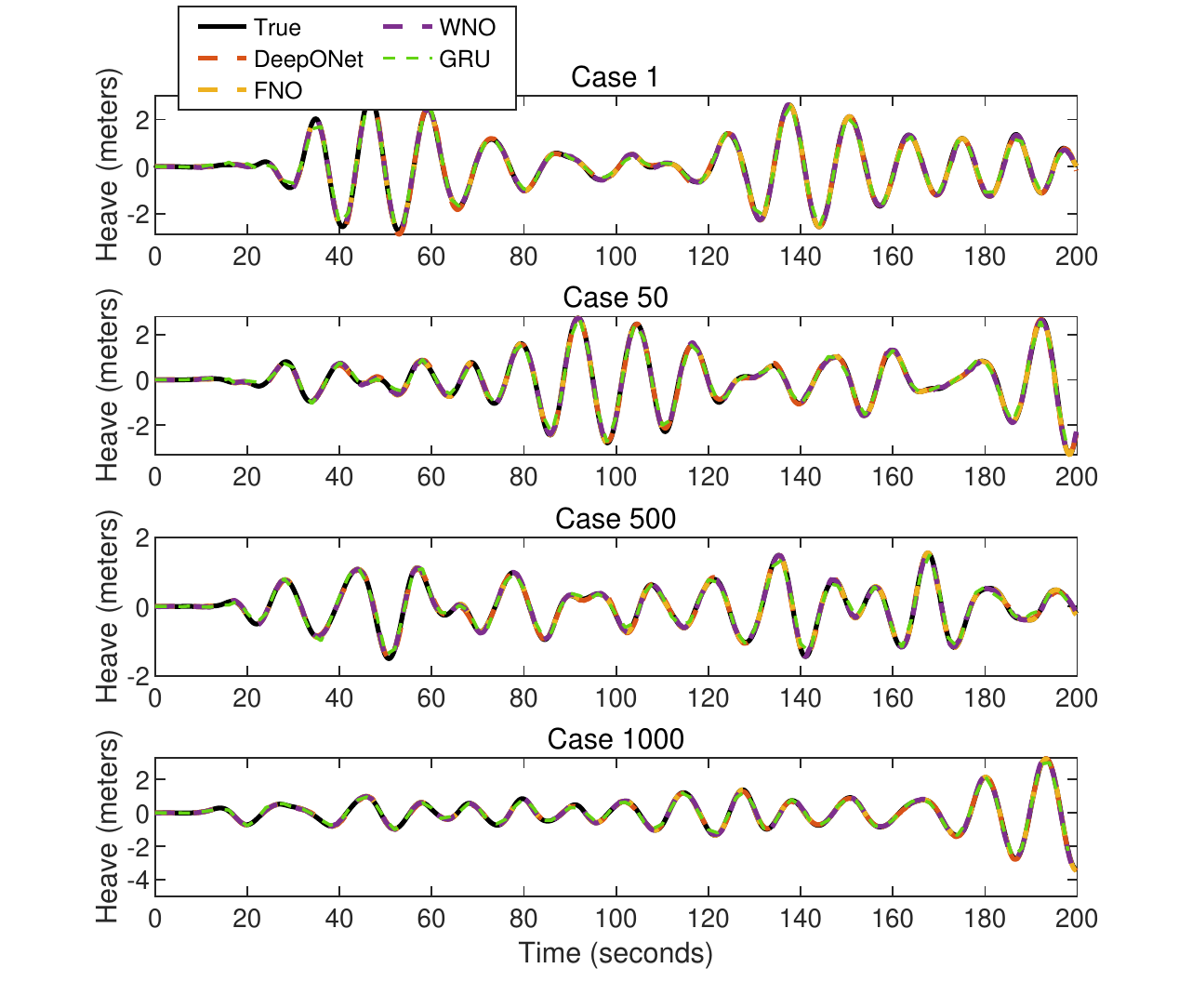}
\end{minipage}
\begin{minipage}[t]{0.5\linewidth}
\centering
\includegraphics[width=3in]{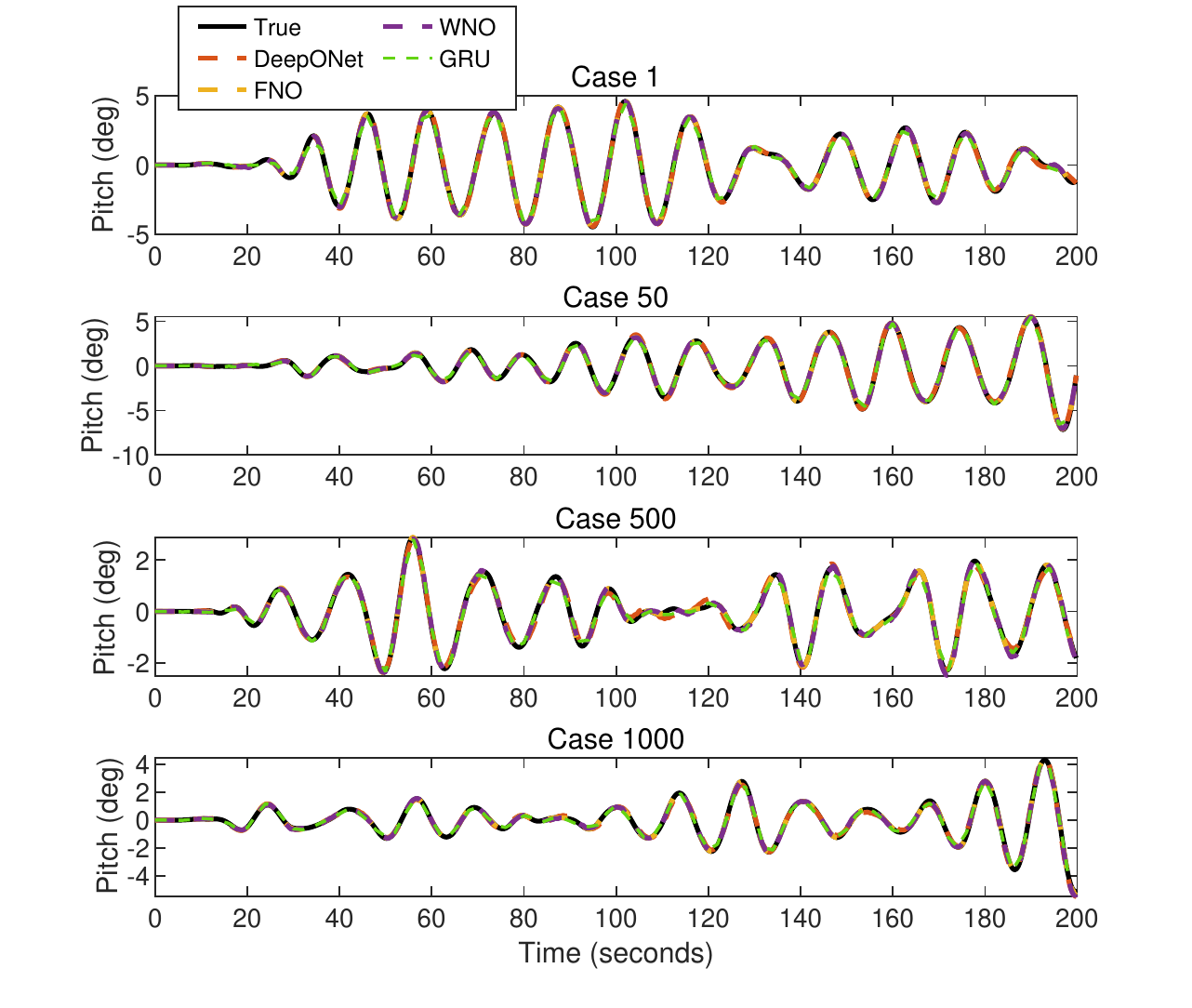}
\end{minipage}
\centering
\caption{Comparison of heave and pitch responses predicted by the four models, DeepONet, FNO, WNO and GRU against the ground truth in Case 2. The plots on left show the heave motion, while the plots on right show the pitch motion.}\label{8state}
\end{figure}

\begin{figure}[H]
\centering
\begin{minipage}[t]{0.5\linewidth}
\centering
\includegraphics[width=3in]{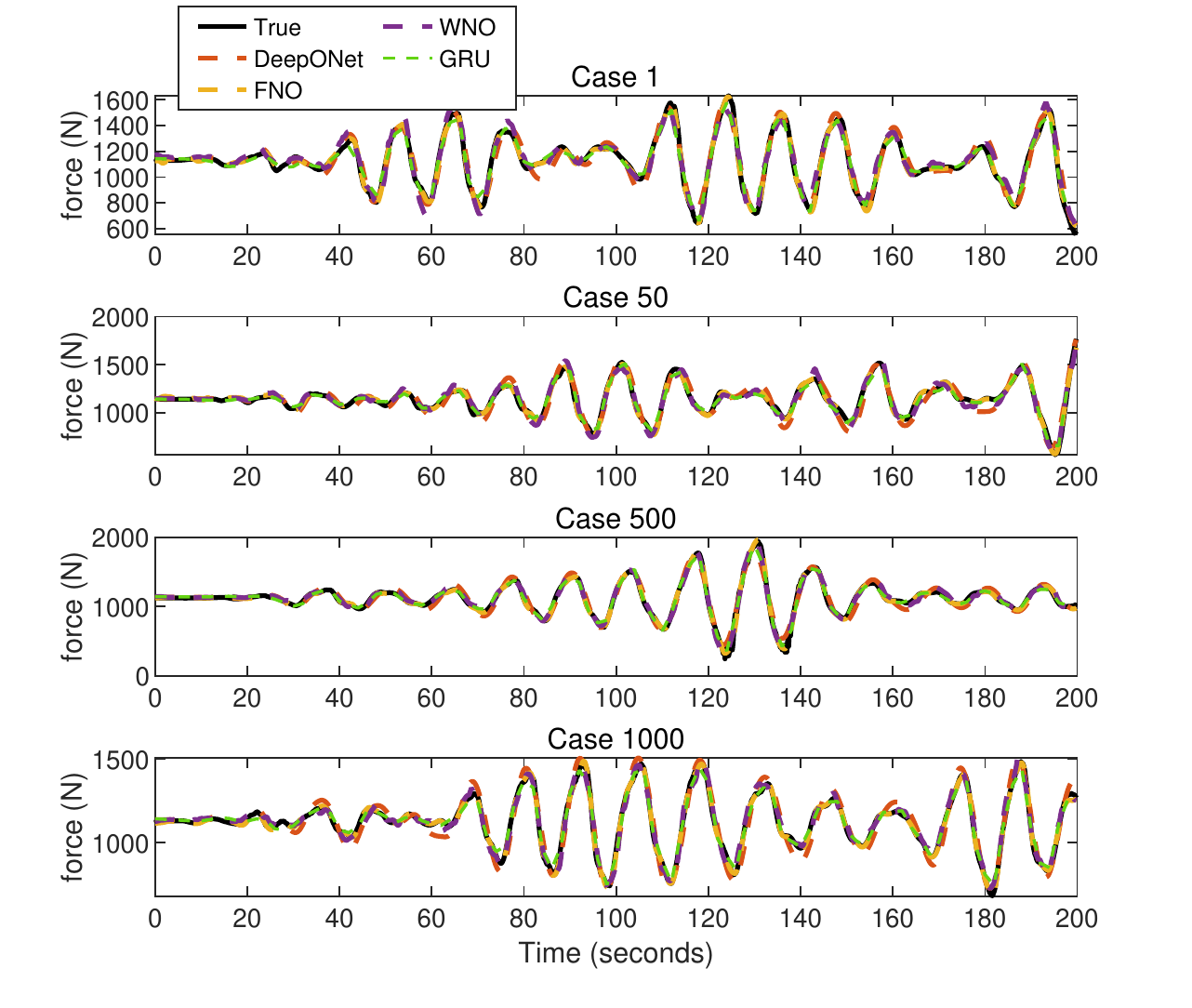}
\end{minipage}
\begin{minipage}[t]{0.5\linewidth}
\centering
\includegraphics[width=3in]{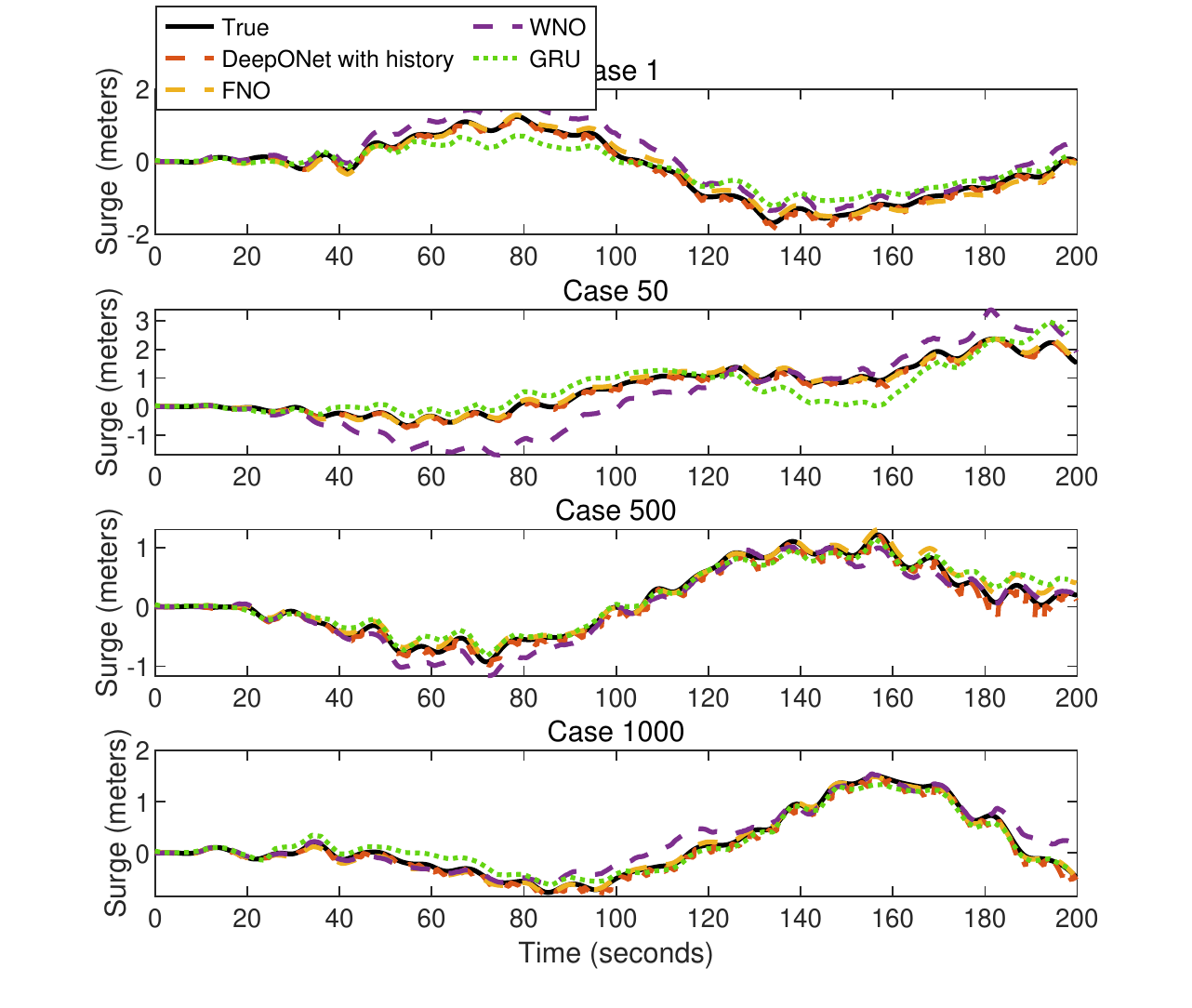}
\end{minipage}
\centering
\caption{Comparison of the top tension of mooring \#1 and surge response predictions obtained by the four models against the ground truth in Case 2. The plots on left show the top tension, while the plots on right show the surge motion.}\label{moor_8state}
\end{figure}

\begin{figure}[H]
\centering
\begin{minipage}[t]{0.5\linewidth}
\centering
\includegraphics[width=3in]{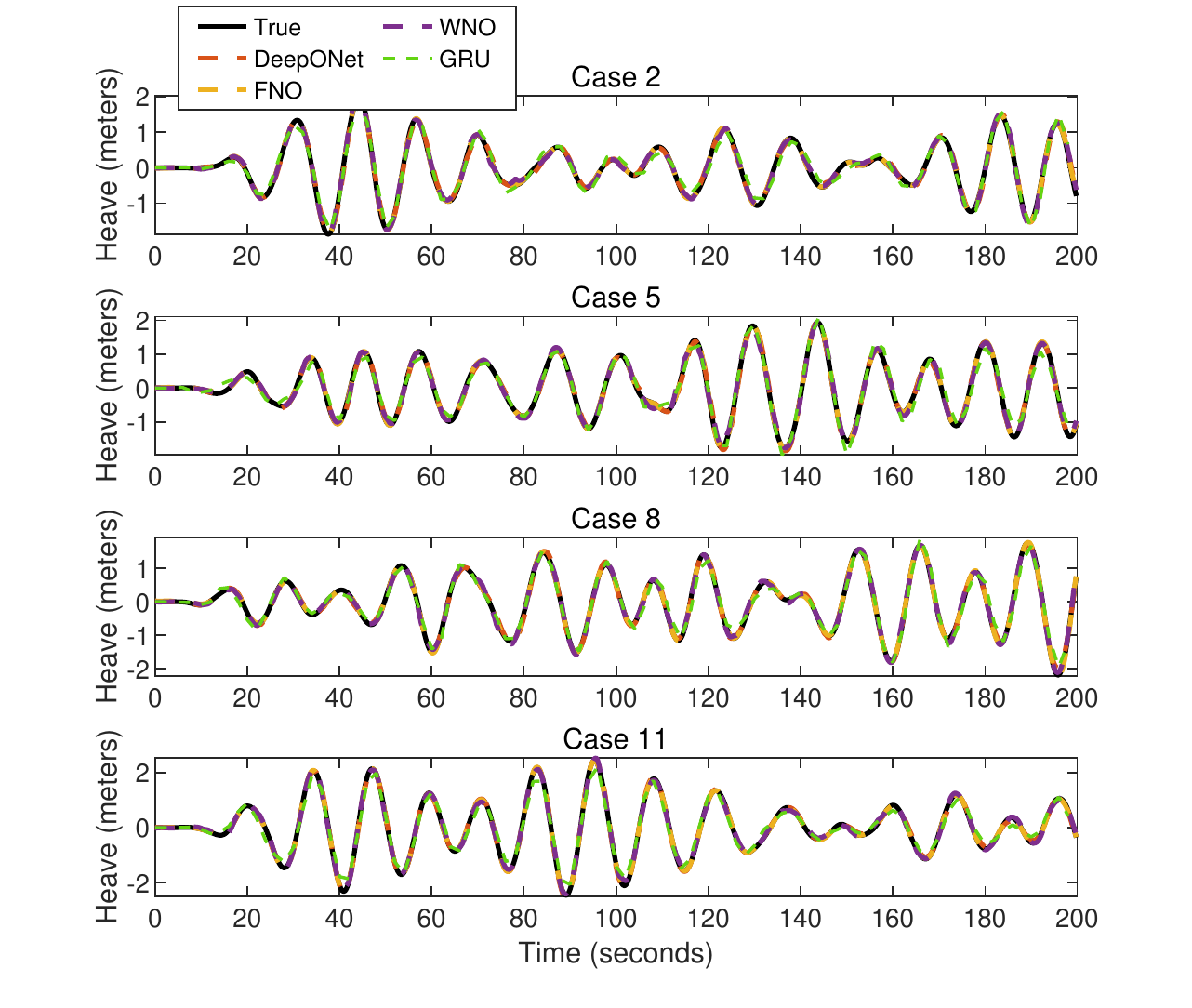}
\end{minipage}
\begin{minipage}[t]{0.5\linewidth}
\centering
\includegraphics[width=3in]{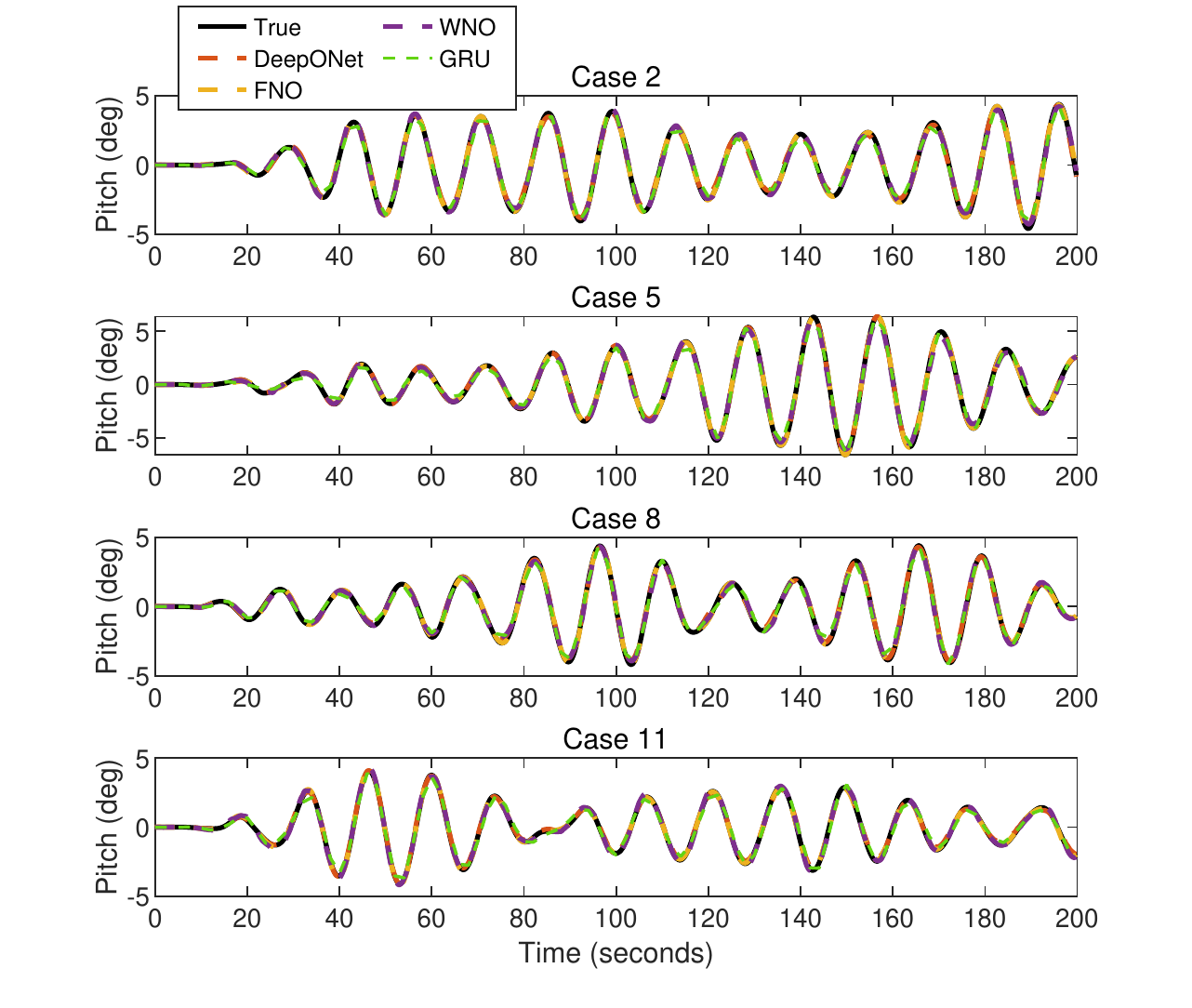}
\end{minipage}
\centering
\caption{Comparison of heave and pitch responses predicted by the four models, DeepONet, FNO, WNO and GRU against the ground truth in Case 3. The plots on left show the heave motion, while the plots on right show the pitch motion.}\label{allstate}
\end{figure}

\begin{figure}[H]
\centering
\begin{minipage}[t]{0.5\linewidth}
\centering
\includegraphics[width=3in]{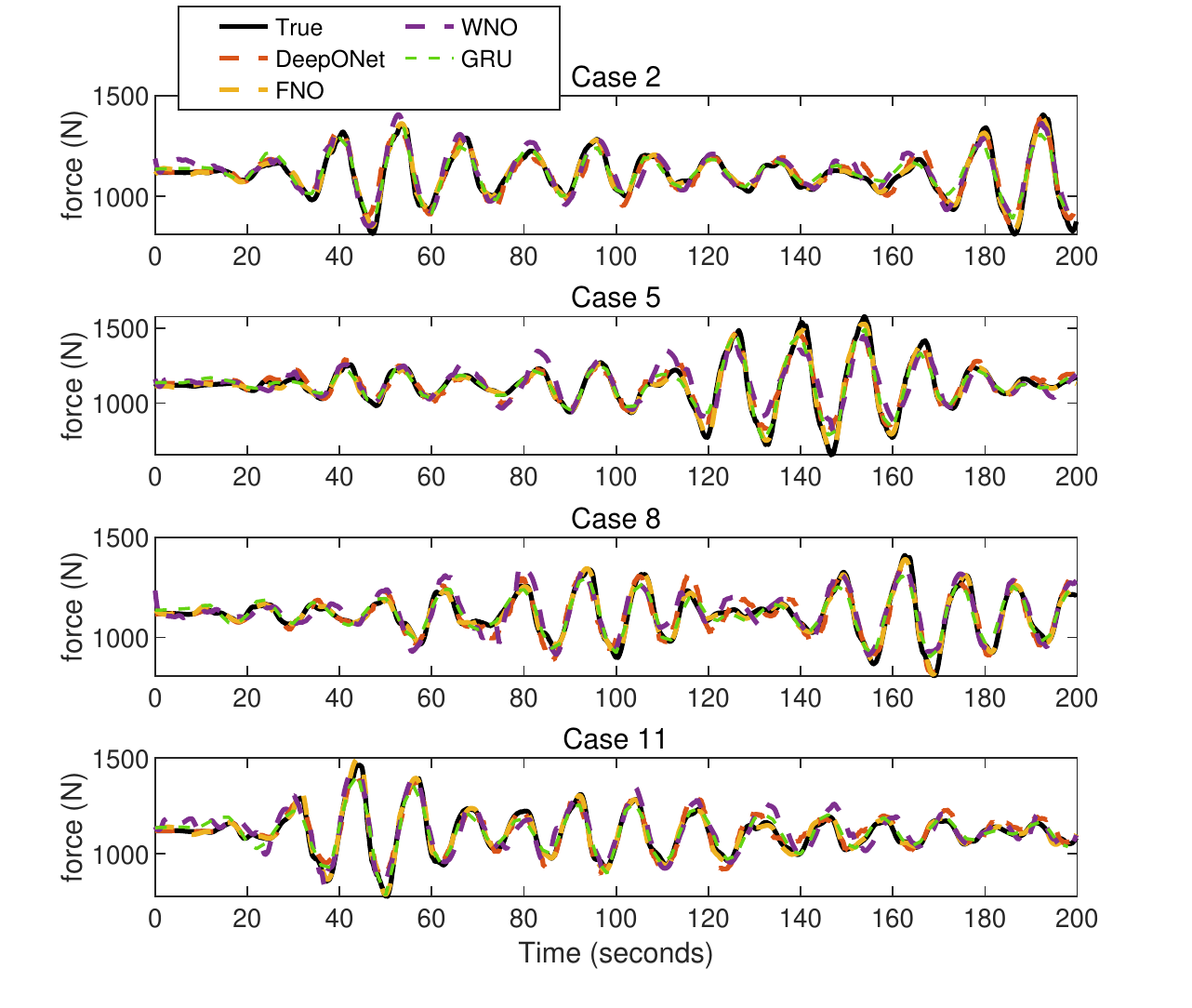}
\end{minipage}
\begin{minipage}[t]{0.5\linewidth}
\centering
\includegraphics[width=3in]{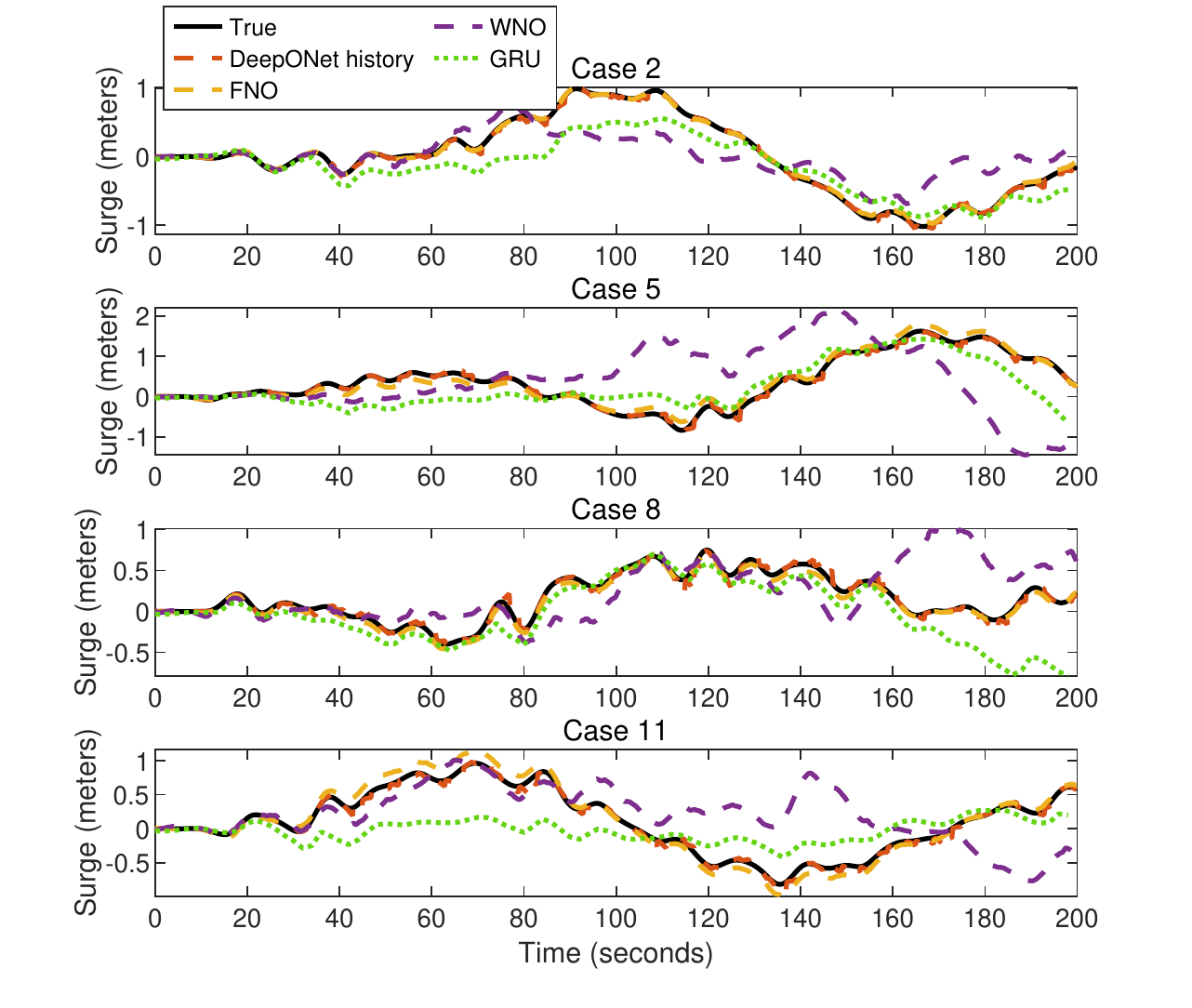}
\end{minipage}
\centering
\caption{Comparison of the top tension of mooring \#1 and surge response predictions obtained by the four models against the ground truth in Case 3. The plots on the left show the top tension, while the plots on the right show the surge motion.}\label{moor_allstate1}
\end{figure}

\begin{figure}[H]
\centering
\begin{minipage}[t]{0.5\linewidth}
\centering
\includegraphics[width=3in]{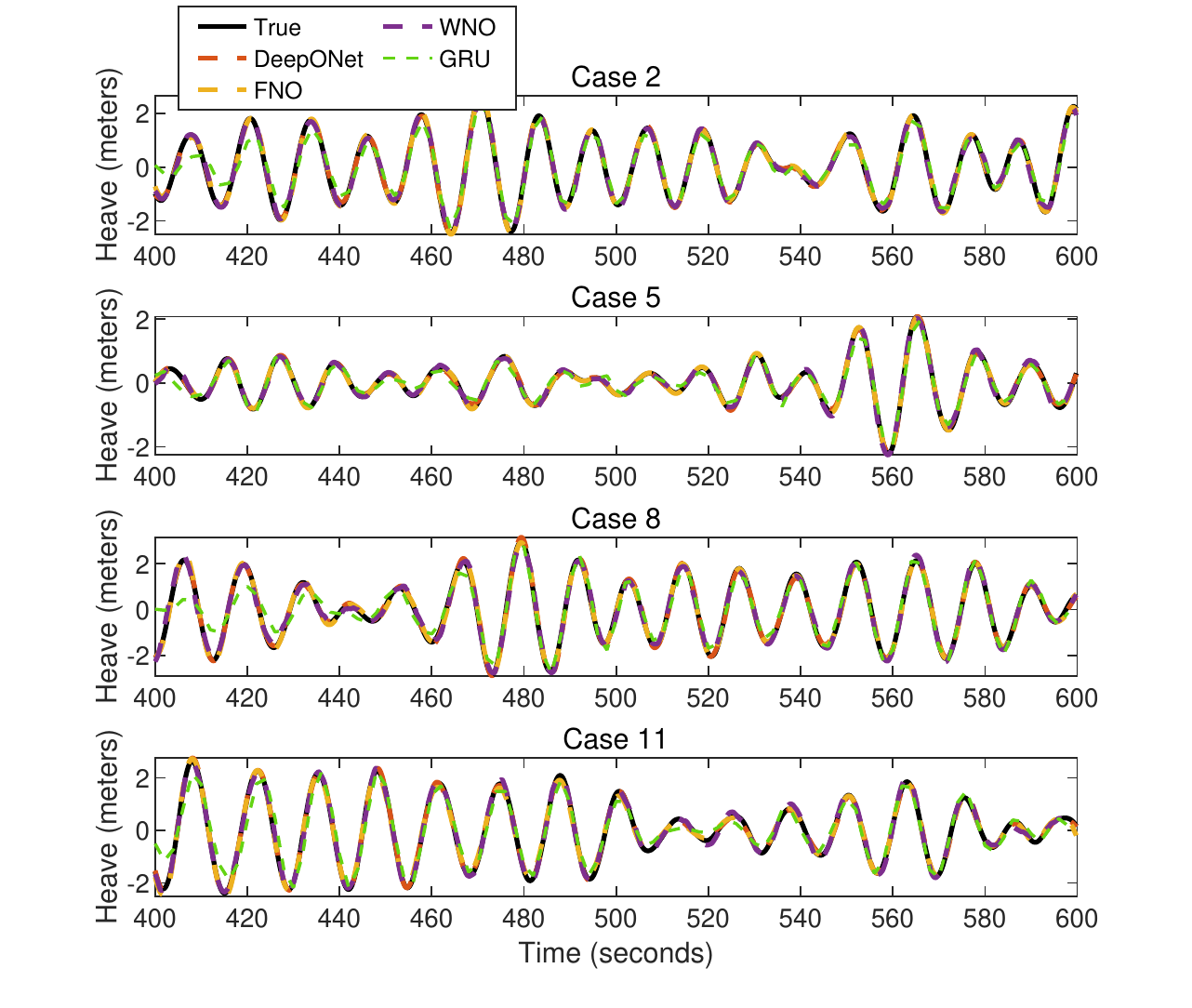}
\end{minipage}
\begin{minipage}[t]{0.5\linewidth}
\centering
\includegraphics[width=3in]{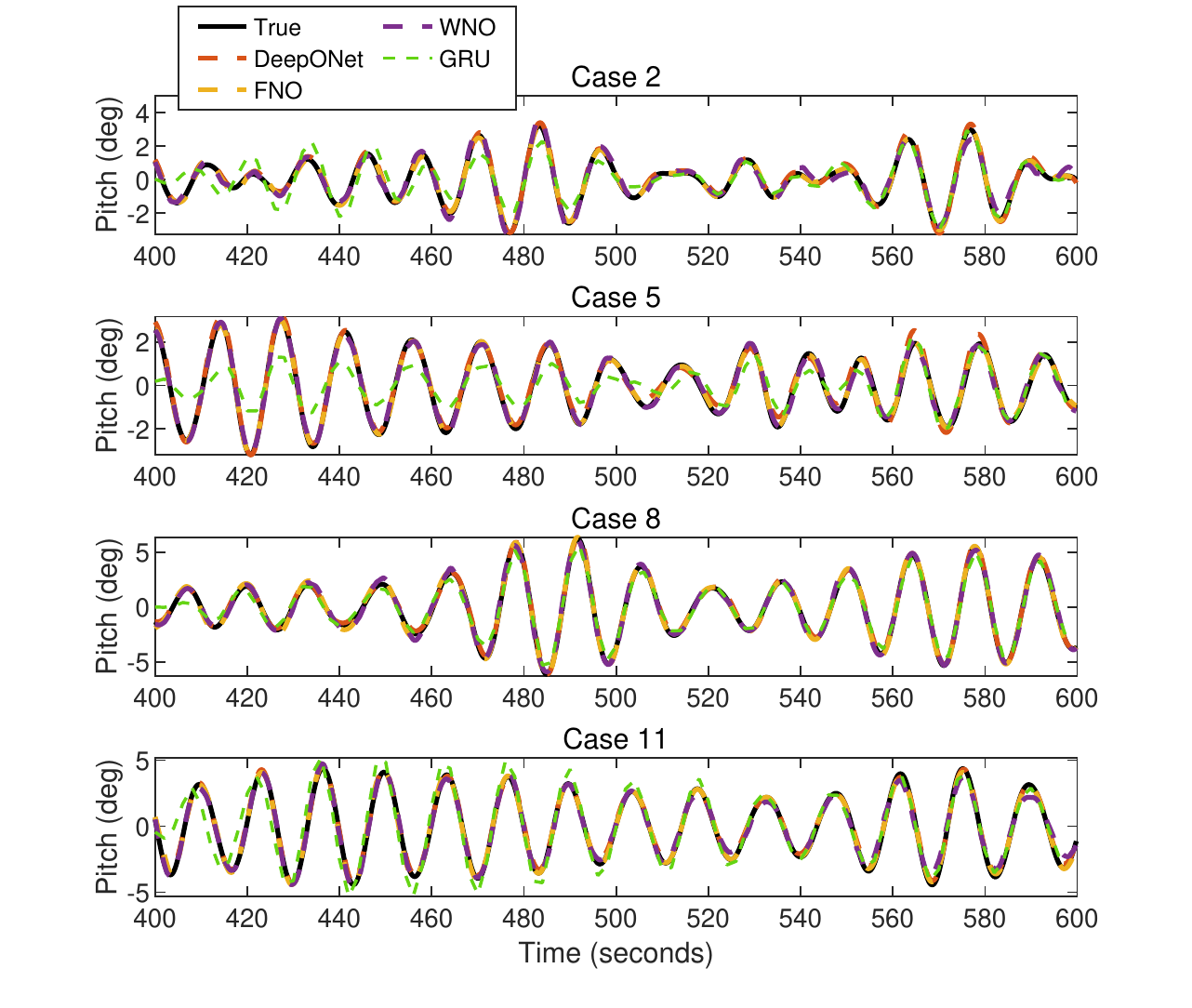}
\end{minipage}
\centering
\caption{Comparison of heave and pitch responses predicted by the four models, DeepONet, FNO, WNO, and GRU against the ground truth in Case 8. The plots on the left show the heave motion, while the plots on the right show the pitch motion.}\label{allstate2}
\end{figure}

\begin{figure}[H]
\centering
\begin{minipage}[t]{0.5\linewidth}
\centering
\includegraphics[width=3in]{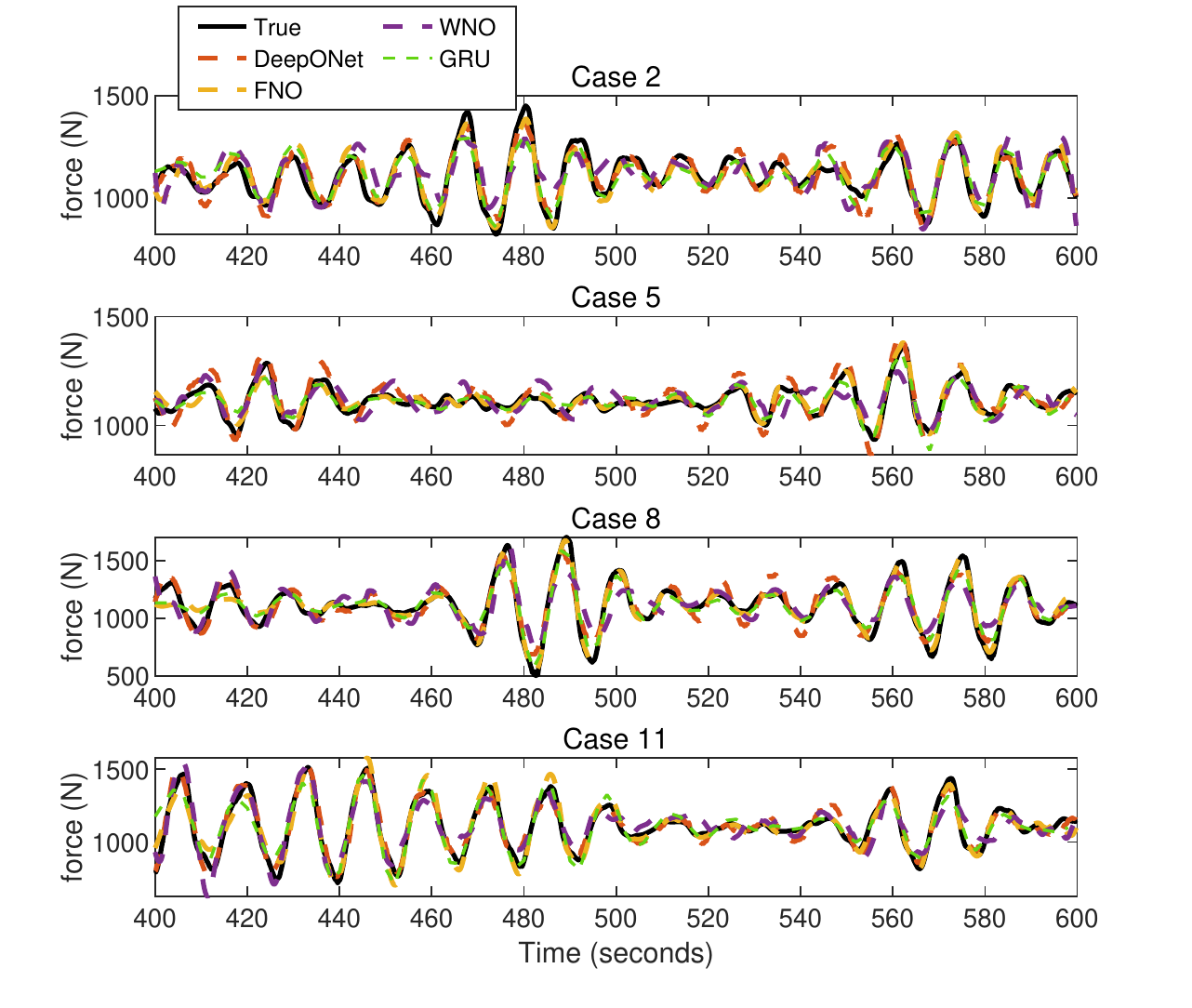}
\end{minipage}
\begin{minipage}[t]{0.5\linewidth}
\centering
\includegraphics[width=3in]{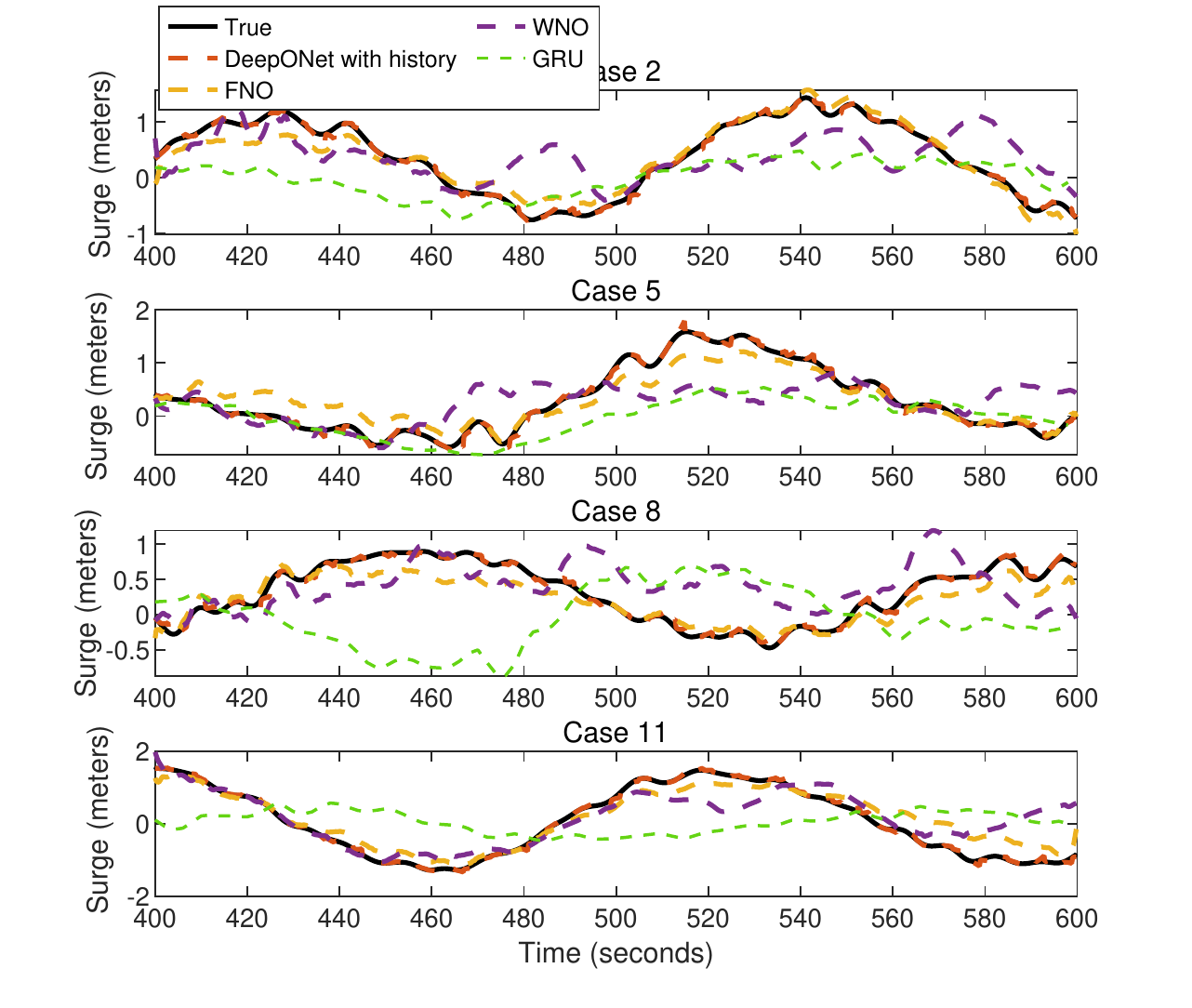}
\end{minipage}
\centering
\caption{Comparison of the top tension of mooring \#1 and surge response predictions obtained by the four models against the ground truth in Case 8. The plots on the left show the top tension, while the plots on the right show the surge motion.}\label{moor_allstate2_1}
\end{figure}

This section gives some representative plots of the response from neural operators. Figs.~\ref{1state}-\ref{moor_1state} show the comparison of heave, pitch, top tension of mooring line and surge predicted by the four models, DeepONet, FNO, WNO, and GRU against the ground truth in Case 1. Figs.~\ref{8state}-\ref{moor_8state} show the comparison of heave, pitch, top tension of mooring line and surge predicted by the four models, DeepONet, FNO, WNO, and GRU against the ground truth in Case 2. Figs.~\ref{allstate}-\ref{moor_allstate1} show the comparison of heave, pitch, top tension of mooring line and surge predicted by the four models, DeepONet, FNO, WNO, and GRU against the ground truth in Case 3. Figs.~\ref{allstate2}-\ref{moor_allstate2_1} show the comparison of heave, pitch, top tension of mooring line and surge predicted by the four models, DeepONet, FNO, WNO, and GRU against the ground truth in Case 8.

\section{Mean and standard deviation of MSE}
\label{app:mean_std_error}
To evaluate the performance of the model, we compute the mean square error of the predictions, and we report the mean and standard deviation of this metric based on three independent training trials in Table ~\ref{MSE_surge_non_zero_initial}.

\begin{table}[H]
\renewcommand\arraystretch{1.15}
\centering
\caption{Mean Square error of predicted surge by operator networks with different initial conditions}\label{MSE_surge_non_zero_initial}
\scriptsize
\begin{tabular}{lcccccccccc}
\hline
Methods  & \multicolumn{2}{c}{Case 1}   & \multicolumn{2}{c}{Case 2}  & \multicolumn{2}{c}{Case 3} & \multicolumn{2}{c}{Case 4} &  \multicolumn{2}{c}{Case 5}\\
\hline
    & Mean & Std & Mean & Std & Mean & Std & Mean & Std & Mean & Std\\
Vanilla DeepONet   &2.87e-2 & 3.04e-3 &3.16e-2 & 3.27e-3  &2.99e-1 & 4.73e-3 & 1.14e-1& 6.66e-3 &9.91e-2 & 5.06e-3\\
POD-DeepONet& 1.50e-2 & 1.99e-3&3.26e-2  &4.57e-3 &1.32e-1  &8.08e-3 &7.38e-2  &2.32e-2 &3.58e-2  &3.82e-3\\
DeepONet with history & 4.43e-4 & 2.43e-4& 2.00e-3 &8.56e-4 &  1.55e-3 &9.07e-5 &9.91e-4 &3.76e-5 &7.32e-4    &1.19e-4\\
SA-DeepONet& 2.00e-2 & 4.93e-4&2.80e-2  &1.94e-2 &7.43e-2  &1.13e-2 & 9.18e-3 & 7.16e-4& 8.34e-3 &1.74e-3\\
W-DeepONet   &2.43e-2  &1.76e-3 &8.43e-3  &5.15e-4 &2.91e-1  &3.59e-2 & 8.16e-3 &9.15e-4 & 6.07e-3 &6.50e-4\\
FNO   & 3.39e-3 & 9.54e-5 &5.03e-4  &4.79e-5 & 9.36e-3 & 1.31e-3& 3.11e-3 & 2.17e-4&1.93e-3  &5.13e-5\\
WNO  & 3.65e-2 &1.17e-3&1.07e-1  & 4.00e-3&1.81e-1  &1.73e-2 &2.36e-2  &4.36e-4 &1.68e-2  &7.09e-4\\
SA-WNO  &1.49e-2  &1.53e-4 & 1.03e-2 &2.06e-3 &3.97e-2 &1.13e-3  &8.83e-3  & 7.45e-4& 8.06e-3 &5.12e-4\\
GRU  & 2.69e-3 &4.04e-5 &7.42e-2  &2.65e-4 &4.24e-1  &1.15e-3 & 1.73e-1 &1.15e-3 & 5.07e-2 &2.08e-4\\
\hline
 & \multicolumn{2}{c}{Case 6}   & \multicolumn{2}{c}{Case 7}  & \multicolumn{2}{c}{Case 8} & \multicolumn{2}{c}{Case 9} &  \multicolumn{2}{c}{Case 10}\\
\hline
    & Mean & Std & Mean & Std & Mean & Std & Mean & Std & Mean & Std\\
DeepONet history & 3.22e-3 & 4.89e-4& 6.31e-2 & 1.50e-3& 1.80e-3 &5.89e-4 & 1.27e-3 & 1.32e-4& 1.22e-3 &2.52e-5\\
W-DeepONet   & 4.15e-2 &3.03e-3& 6.55e-2 &6.18e-3& 3.43e-1 &2.46e-2& 5.92e-2 &1.31e-3& 3.75e-2 &5.97e-3\\
FNO  & 3.14e-2 &5.60e-3& 7.72e-2 &5.22e-3& 9.55e-2 &4.73e-4&3.98e-2 &4.88e-4& 1.08e-3&2.98e-4\\
WNO  & 1.39e-1 &1.04e-2&  4.40e-1 &2.52e-3& 3.74e-1 &6.43e-3& 1.35e-1 &2.15e-2& 1.23e-1&1.74e-2\\
SA-WNO & 6.22e-2 &6.21e-3& 1.05e-1 &1.27e-2& 3.30e-1 &1.42e-2& 5.87e-2 &4.60e-3& 4.98e-2 &3.61e-3\\
GRU & 1.57e-1 &7.64e-3& 4.29e-1 &1.74e-2& 4.23e-1 &6.00e-3& 3.57e-1 &3.06e-3& 2.99e-1 &4.62e-2\\
\hline
\end{tabular}
\end{table}

\section{Network architectures}
\label{app:architectures}
This section gives the network architectures used in different models for training an operator. Tables ~\ref{DeepONet_zero}-\ref{GRU_zero} summarize hyperparameters used in the DeepONet, FNO, WNO, GRU for training an operator to approximate heave, pitch and tension. Tables ~\ref{parameters_surge_1seastate_zero_initial2}-\ref{parameters_surge_1seastate_zero_initial1} summarize hyperparameters used in the model for training an operator to approximate the surge for considering zero initial conditions of displacement and velocities. Tables ~\ref{parameters_surge_1seastate_zero_initial11}-\ref{parameters_surge_1seastate_zero_initial} summarize hyperparameters used in the model for training an operator to approximate the surge for considering different initial conditions of displacement and velocities.
\begin{table}[H]
\renewcommand\arraystretch{1.15}
\centering
   \caption{Hyperparameters used in the DeepONet model for training an operator to approximate the degrees of freedom considering zero initial conditions of displacement and velocities.}
   \label{DeepONet_zero}
   \scriptsize
   \begin{tabular}{ccccccccc}
   \hline
     Sea state & Response & Model & NN depth  & NN width  & Modes & Activation function & Learning rate & Iterations \\
   \hline
   \multirow{3}{*}{All cases}    & \multirow{3}{*}{Heave} & \multirow{3}{*}{vanilla DeepONet} & \multirow{3}{*}{3}  & \multirow{3}{*}{100}    & \multirow{3}{*}{$/$}  & \multirow{3}{*}{sin} & \multirow{3}{*}{1e-4}   & \multirow{3}{*}{100,000}\\
    &  & & & & &&\\
   &  & & & & &&\\
\hline
   \multirow{3}{*}{All cases}    & \multirow{3}{*}{Pitch} & \multirow{3}{*}{vanilla DeepONet} & \multirow{3}{*}{3}  & \multirow{3}{*}{100}    & \multirow{3}{*}{$/$} & \multirow{3}{*}{sin} & \multirow{3}{*}{1e-4}   & \multirow{3}{*}{100,000}\\
    &  & & & & &&\\
    &  & & & & &&\\
\hline
    \multirow{3}{*}{All cases}  & \multirow{3}{*}{Tension} & \multirow{3}{*}{vanilla DeepONet} & \multirow{3}{*}{4}  & \multirow{3}{*}{256}    & \multirow{3}{*}{$/$} & \multirow{3}{*}{sin} & \multirow{3}{*}{1e-4}   & \multirow{3}{*}{400,000}\\
    &  & & & & &&\\
    &  & & & & &&\\
\hline
   \end{tabular}
\end{table}

\begin{table}[H]
\centering
   \caption{Hyperparameters used in the FNO for training an operator to approximate the degrees of freedom considering zero initial conditions of displacement and velocities.}
   \label{FNO_zero}
   \scriptsize
   \begin{tabular}{cccccccc}
   \hline
     Sea states & Response      & Fourier layer  & NN width  & Modes & Activation function & Learning rate & Epochs \\
   \hline
   \multirow{3}{*}{All cases}    & Heave     & \multirow{3}{*}{4}  &  \multirow{3}{*}{32}    & \multirow{3}{*}{50}  &\multirow{3}{*}{ReLU} &\multirow{3}{*}{1e-3}   & \multirow{3}{*}{300} \\
    & Pitch & & & & &\\
    & Tension & & & & &\\
   \hline
   \end{tabular}
\end{table}

\begin{table}[H]
\renewcommand\arraystretch{1.15}
\centering
   \caption{Hyperparameters used in the WNO for training an operator to approximate the degrees of freedom considering zero initial conditions of displacement and velocities.}
   \label{WNO_zero}
   \scriptsize
   \begin{tabular}{ccccccccc}
   \hline
      Sea states & Response      &   Mother Wavelet  & Level & Layer  & NN width  &Activation function & Learning rate & Epochs \\
   \hline
    \multirow{3}{*}{All cases} &Heave     & \multirow{3}{*}{ db20} &\multirow{3}{*}{6}   & \multirow{3}{*}{4}  &  \multirow{3}{*}{64}   &\multirow{3}{*}{ReLU}   &\multirow{3}{*}{1e-3}   & \multirow{3}{*}{300}\\
       & Pitch & & & & &\\
    & Tension & & & & &\\
    \hline
   \end{tabular}
\end{table}

\begin{table}[H]
\centering
   \caption{Hyperparameters of the GRU with zero initial conditions}
   \label{GRU_zero}
   \scriptsize
   \begin{tabular}{cccccccc}
   \hline
     Sea state & Response  & Number of hidden layer  & NN width & Number of dense layer  & Learning rate & Iterations \\
   \hline
   \multirow{3}{*}{Case 1} & Heave   &  \multirow{3}{*}{1}  &   \multirow{3}{*}{50}    &  \multirow{3}{*}{1} &    \multirow{3}{*}{1e-4}   &  \multirow{3}{*}{3,000}\\

   & Pitch   &   &     &  &   &    \\

   & Tension  &   &     &  &   &    \\

   \hline
      \multirow{3}{*}{Case 2} & Heave   &  \multirow{3}{*}{2}  &   \multirow{3}{*}{200}    &  \multirow{3}{*}{1} &    \multirow{3}{*}{1e-4}   &  \multirow{3}{*}{3,000}\\

   & Pitch   &   &     &  &   &    \\

   & Tension  &   &     &  &   &    \\

   \hline
      \multirow{3}{*}{Case 3} & Heave   &  \multirow{3}{*}{2}  &   \multirow{3}{*}{200}    &  \multirow{3}{*}{1} &    \multirow{3}{*}{1e-5}   &  \multirow{3}{*}{1,000}\\

   & Pitch   &   &     &  &   &    \\

   & Tension  &   &     &  &   &    \\

   \hline
   \end{tabular}
\end{table}

\begin{table}[H]
\renewcommand\arraystretch{1.15}
\centering
  \caption{Hyperparameters used in the model for training a DeepONet to approximate the surge for considering zero initial conditions of displacement and velocities.}\label{parameters_surge_1seastate_zero_initial2}
  \scriptsize
  \begin{tabular}{lcccccc}
  \hline
\multirow{2}{4em}{Model} & \multirow{2}{6em}{Branch net}    & \multirow{2}{6em}{Trunk net} & Activation & Learning & \multirow{2}{4em}{Iterations} & \multirow{2}{4em}{Others}\\
 & & & & function & rate &\\
 \hline
  Vanilla DeepONet & [750, 750, 750, 750] & [750, 750, 750, 750] & sin & 1e-4 & 100,000 & $/$\\
  POD-DeepONet  & [750, 750, 750,100] &  [750, 750, 750,100]& sin &vary lr & 1,000,000 & $/$\\
  DeepONet with history  &[100,100,300] & [100,100,300] & sin  & vary lr& 100,000 & 5\\
  SA-DeepONet   & [512, 512, 512, 512] & [512, 512, 512, 512] & sin  & vary lr& 1,000,000 & $/$\\
  W-DeepONet   & [128, 128, 128, 300] &  [128, 128, 300] & sin & vary lr & 1,000,000 &db1(Mother wavelet)\\
  \hline
  \end{tabular}
\end{table}

\begin{table}[H]
\renewcommand\arraystretch{1.15}
\centering
  \caption{Hyperparameters used in the model for training an operator to approximate the surge for considering zero initial conditions of displacement and velocities.}\label{parameters_surge_1seastate_zero_initial1}
  \scriptsize
  \begin{tabular}{lccccccc}
  \hline
\multirow{2}{4em}{Model} & \multirow{2}{4em}{Depth}  & \multirow{2}{4em}{Width}  & \multirow{2}{4em}{Modes} & Activation & Learning & \multirow{2}{4em}{Epochs} & \multirow{2}{4em}{Others}\\
 & & & & function & rate &\\
 \hline
  FNO  & 4 & 32 & 501 & ReLU & 1e-3 & 300 &$/$\\
  WNO  & 4 & 64 & 6 (level of wavelet tree) & ReLU & 1e-3& 300  & db20 (Mother wavelet)\\
  SA-WNO  & 4 & 48 & 6 (level of wavelet tree) & GELU & 1e-3& 300  & db24 (Mother wavelet)\\
  GRU (Case 1)  & 1 & 50 & $/$ &  & 1e-4& 3,000 &$/$\\
  GRU (Case 2)  & 2 & 200 & $/$ &  & 1e-4& 3,000 &$/$\\
  GRU (Case 3)  & 2 & 200 & $/$ &  & 1e-5& 1,000 &$/$\\
GRU (Case 4)  & 2 & 200 & $/$ &  & 1e-5& 1,000 &$/$\\
GRU (Case 5)  & 2 & 200 & $/$ &  & 1e-5& 1,000 &$/$\\
  \hline
  \end{tabular}
\end{table}

\begin{table}[H]
\renewcommand\arraystretch{1.15}
\centering
  \caption{Hyperparameters used in the model for training a DeepONet to approximate the surge for considering different initial conditions of displacement and velocities.}\label{parameters_surge_1seastate_zero_initial11}
  \scriptsize
  \begin{tabular}{lccccccc}
  \hline
\multirow{2}{4em}{Model} & \multirow{2}{6em}{Branch net}   & \multirow{2}{6em}{Trunk net} & Activation & Learning & \multirow{2}{4em}{Iterations} & \multirow{2}{4em}{Others}\\
 & & & function & rate &\\
 \hline
  DeepONet with history (Cases 6/8-10)  &[100,100,300] & [100,100,300] & sin  & vary lr& 100,000 & 5 steps\\
DeepONet with history (Case 7)  &[100,100, 100, 100] & [100,100,100] & sin  & vary lr& 100,000 & 5 steps\\
W-DeepONet(Case 6) &  [128, 128, 128, 128] &  [128, 128, 128] & sin & vary lr & 1,000,000 &db1 (Mother wavelet)\\
W-DeepONet(Cases 7/9) &  [128, 128, 128, 800] &  [128, 128, 800] & sin & vary lr & 1,000,000 &db1 (Mother wavelet)\\
W-DeepONet(Case 8) &  [128, 128, 128, 500] &  [128, 128, 500] & sin & vary lr & 1,000,000 &db1 (Mother wavelet)\\
W-DeepONet(Case 10)   & [256, 256, 256, 800]&   [256, 256, 800] & sin & vary lr & 1,000,000 &db1 (Mother wavelet)\\
  \hline
  \end{tabular}
\end{table}

\begin{table}[H]
\renewcommand\arraystretch{1.15}
\centering
  \caption{Hyperparameters used in the model for training an operator to approximate the surge for considering different initial conditions of displacement and velocities.}\label{parameters_surge_1seastate_zero_initial}
  \scriptsize
  \begin{tabular}{lccccccc}
  \hline
\multirow{2}{4em}{Model} & \multirow{2}{4em}{Depth}   & \multirow{2}{4em}{Width} & \multirow{2}{4em}{Modes} & Activation & Learning & \multirow{2}{4em}{Epochs} & \multirow{2}{4em}{Others}\\
 & & & & function & rate &\\
 \hline
  FNO  & 4 & 32 & 501 & ReLU & 1e-3 & 300 &$/$\\
  WNO  & 4 & 64 & 6 (level of wavelet tree) & ReLU & 1e-3& 300  & db20 (Mother wavelet)\\
SA-WNO  & 4 & 32 & 6 (level of wavelet tree) & GELU & 1e-3& 300  & db24 (Mother wavelet)\\
  GRU (Case 6)  & 1 & 50 & $/$ &  & 1e-4& 3,000 &$/$\\
  GRU (Case 7)  & 2 & 200 & $/$ &  & 1e-4& 3,000 &$/$\\
  GRU (Case 8)  & 2 & 200 & $/$ &  & 1e-5& 1,000 &$/$\\
GRU (Case 9)  & 2 & 200 & $/$ &  & 1e-5& 1,000 &$/$\\
GRU (Case 10)  & 2 & 200 & $/$ &  & 1e-5& 1,000 &$/$\\
  \hline
  \end{tabular}
\end{table}

\section{Training time of network}
\label{app:architectures}
This section shows the training times of the network to approximate the surge for all cases, which are summarized in Table ~\ref{MSE_surge_zero_initial}.
   \begin{table}[H]
\renewcommand\arraystretch{1.5}
\centering
  \caption{The training times of the network to approximate the surge for all cases (seconds)}\label{MSE_surge_zero_initial}
  \tiny
  \begin{tabular}{lcccccccccc}
  \hline
     Methods  & Case 1   &  Case 2  &  Case 3  &  Case 4  &  Case 5  & Case 6   &  Case 7  &  Case 8  &  Case 9  &  Case 10\\
 \hline
  Vanilla DeepONet &651,157&1,376,428&558,278&348,916&419,246&/&/&/&/&/\\
  POD-DeepONet  &23,712&31,750&24,327&20,013&23,781&/&/&/&/&/\\
  DeepONet history &1,571,231&1,388,057&41,823&1,386,504&1,401,539&1,477,188&885,361&48,214&1,471,251&1,550,089\\
  SA-DeepONet  &6,337,139&6,290,858&5,668,849&6,202,152&6,301,864&/&/&/&/&/\\
  W-DeepONet   &627,445&669,708&35,691&640,662&644,568&808,109&1,013,005&53,047&1,655,794&4,835,658\\
  FNO &1,230&2,425&191&2,799&3,765&1,763&4,002&245&4,030&7,252\\
  WNO  &8,055&15,718&1,116&15,699&23,572&8,367&16,850&1,177&17,942&26,102\\
  SA-WNO &7,770&15,455&2,120&16,295&24,758&8,181&14,955&1,791&13,323&25,713\\
  GRU  &4,380&22,712&1,511&23,200&31,891&4,818&20,911&1,490&22901&30,771\\
  \hline
  \end{tabular}
\end{table}

\end{document}